\pdfoutput=1
\documentclass[letterpaper,twocolumn,10pt]{article}

\PassOptionsToPackage{table}{xcolor}
\usepackage{usenix}

\usepackage{graphicx}
\usepackage{amsmath}
\usepackage{amssymb}
\usepackage{amsthm}
\theoremstyle{definition}
\newtheorem{definition}{Definition}
\newtheorem{goal}{Goal}
\theoremstyle{plain}
\newtheorem{theorem}{Theorem}
\usepackage{array}
\usepackage{booktabs}
\usepackage{tikz}
\usetikzlibrary{arrows.meta,positioning,fit,shapes.geometric,calc,backgrounds}
\definecolor{tgreen}{RGB}{0,158,60}
\definecolor{tred}{RGB}{214,39,40}
\newcommand{\tctx}{{\color{tgreen}\mathsf{tctx}}}
\newcommand{\uctx}{{\color{tred}\mathsf{uctx}}}

\newcommand{\rsrc}{{\color{tred}\mathsf{rsrc}}}
\usepackage{xspace}
\usepackage{xurl}
\usepackage{listings}
\usepackage[capitalise,nameinlink]{cleveref}
\crefname{section}{Section}{Sections}
\Crefname{section}{Section}{Sections}
\crefname{subsection}{Section}{Sections}
\Crefname{subsection}{Section}{Sections}
\crefname{subsubsection}{Section}{Sections}
\Crefname{subsubsection}{Section}{Sections}
\crefalias{subsection}{section}
\crefalias{subsubsection}{section}
\crefname{figure}{Figure}{Figures}
\Crefname{figure}{Figure}{Figures}
\crefname{table}{Table}{Tables}
\Crefname{table}{Table}{Tables}
\crefname{equation}{Equation}{Equations}
\Crefname{equation}{Equation}{Equations}
\crefname{lstlisting}{Listing}{Listings}
\Crefname{lstlisting}{Listing}{Listings}
\crefname{definition}{Definition}{Definitions}
\crefname{goal}{Goal}{Goals}
\crefname{theorem}{Theorem}{Theorems}
\Crefname{theorem}{Theorem}{Theorems}
\Crefname{goal}{Goal}{Goals}
\Crefname{definition}{Definition}{Definitions}
\crefname{appendix}{Appendix}{Appendices}
\Crefname{appendix}{Appendix}{Appendices}

\newcommand{\sys}{Pincer\xspace}

\definecolor{syscolor}{gray}{0.94}
\newcolumntype{G}{>{\columncolor{syscolor}}c}

\newcommand{\thead}[1]{\textbf{#1}}
\newcommand{\tsub}[1]{\textcolor{black!55}{#1}}

\begin{document}

\date{}

\title{\Large \bf \sys: Resource Authorization for Agents using a Digital Twin}

\author{
{\rm Mayank Rathee$^{1}$ \qquad Alexander Stepanov$^{1}$ \qquad Shalin Madabhavi$^{2}$}\\
{\rm Jinhao Zhu$^{1}$ \qquad Raluca Ada Popa$^{1}$ \qquad Ion Stoica$^{1}$}\\
{\rm $^{1}$UC Berkeley \qquad $^{2}$Foothill High School}
}

\maketitle

\begin{abstract}
Coding agents have become increasingly long-horizon, autonomous, reliant on general-purpose shell and maintain their own
persistent memory for self-improvement. While these capabilities have made the agents powerful, they have also made them harder to defend against external adversaries. Defenses that restrict
this architecture---typed tools, information-flow control, or policy prediction engines---give up too much functionality to be adopted. Agents deployed today (e.g. Claude, Codex) rely on a combination of user-mediated and auto-mode sandboxing as their primary defense. In user-mediated sandboxing, user-maintained policies decay over time and repeated permission
requests cause user fatigue, while auto mode's tool-call classifiers learn
no user-specific policy and are not meant to defend against adversarial setups.

\sys is a new defense that operates at the resource layer and works alongside existing defenses at the tool-call layer like the auto mode. At the core of \sys lies a \emph{digital
twin}, an isolated-context model that automatically learns and enforces dynamic user-specific least-privilege policies. The digital twin keeps continually learning the user's preferences allowing it to act as the user's proxy for the agent's permission requests.

To emulate the learning phase, we propose a new user-centric dataset with examples following a multi-day transcript of user-agent interaction. Our evaluation shows that~\sys performs strongly on both security and utility in comparison to several baselines which includes variants of LLM judges and adaptations of Conseca (HotOS '25). We highlight attack types where~\sys's design leads to a significant security improvement compared to all other baselines, while outperforming the baselines even for other types of attacks.

\end{abstract}

\section{Introduction}
\label{sec:intro}

Coding agents~\cite{claudecode,codex,openclaw,nanoclaw} have become increasingly autonomous, memory-driven, and reliant on general purpose computer-use tools like shell. These capabilities have ushered in a new era of computing with powerful automation, while also opening the door for vulnerabilities and attackers~\cite{openai2026huggingface,anthropic2026incidents,guardian2026database,howell2026agent,cisco2026memory}. Researchers in both academia and industry have been looking for ways to make agents safer while balancing their usefulness.

In academia, there has been a major push towards strong defenses that in turn heavily restrict the agent's capabilities~\cite{li2026ace,debenedetti2025camel,costa2025fides,gong2026planguard,piet2026planthenexecute,shi2025progent,tsai2025conseca}. On the other hand, the focus in industry has been towards defenses that work well for most day-to-day use and preserving the functionality of agents as much as possible~\cite{anthropic2026automode,anthropic2026automodeconfig,nanoclaw,codex,claudecode}. Sandboxed execution with policies over the agent's privileges, user-mediated permissions, and tool-call classifiers define these defenses. However, their effectiveness tends to decay over time (privilege creep and permission fatigue) and deviation from production traffic, e.g. when an adversary wants to take over the agent, can cause further failures~\cite{ji2026permissiongate}. In this work, we focus primarily on deployed agents and their existing defenses.

We present a new defense that plugs directly into today's agents and targets the resource layer that sits behind tools; for example files, web, etc. Central to our design is an entity we call the {\em digital twin} that learns user behavior over time to automatically infer which resources should be accessible to the agent for any given task. It additionally provides defense-in-depth by staying compatible with existing defenses that are already deployed. We call the system that composes an off-the-shelf agent with our defense {\em\sys}\footnote{Since pincers offer finer control than claws, we use~\sys as our system name to differentiate from less safe agents like OpenClaw.}. Our contributions are as follows:
\begin{enumerate}
    \item \sys, an agentic system that learns user behavior and automatically authorizes resources for the agent on a per task basis.
    \item A user-centric synthetic dataset containing a coherent set of examples from different users' interaction with their agents over several days. The examples contain both attacks and benign workflows. To the best of our knowledge, it is the first dataset of this kind.
    \item Implementation and evaluation of~\sys on top of a lightweight off-the-shelf agent NanoClaw~\cite{nanoclaw}.
\end{enumerate}

\subsection{Technical Overview}

Deployed agents today run in a sandbox where a policy file, maintained by the user, describes what privileges the agent gets within the sandbox including tools it can use, whitelisted commands, and resources it can access. Whenever the agent hits the wall of the sandbox, a permission request is issued to the user specific to where the agent is stuck. Given this user-mediated sandboxing, two challenges show up. First, policies can become over-permissive over time as permissions accumulate because users rarely revoke permissions they might have granted under a deadline; this is known as privilege creep~\cite{felt2011demystified,anthropic2026automodeconfig}. Secondly, repeated permission requests lead to permission fatigue for the users causing their scrutiny to decrease over time~\cite{felt2012attention,li2026conleash}. In response to these concerns, Anthropic released the auto mode~\cite{anthropic2026automode} which is a classifier that sits at the tool-call layer and resolves permission requests to reduce user fatigue. While it works well for regular usage, recent work on stress-testing found that it allows a majority of unsafe operations to go through~\cite{ji2026permissiongate}.
For example, directly executing a dangerous command like \texttt{rm -rf *} is blocked by the classifier, but the same command can be embedded in a harmlessly named Python file \texttt{test.py}, and the classifier allows \texttt{python3 test.py} to run. This gives an adversary an easy way to bypass or fool the classifier. In fact, it is not practical for any tool-call layer classifier to reason through the effects of code, especially when the dangerous command is buried deep in the dependency chain. The same holds even without code: reasoning through the possible effects of a complex shell command at the tool-call layer can be infeasible.
Moreover, the classifier doesn't learn user-specific policies; if the classifier thinks \texttt{git push} is dangerous while the user's workflow deems it as no risk, the user has to manually edit the policy file, otherwise the classifier will keep blocking it. This is why the problem of privilege creep is left unresolved by auto mode. We address these challenges in~\sys and describe our design next.

\paragraph{Privilege creep and permission fatigue.} The primary cause of privilege creep is policies that don't adapt over time, go stale, and become over-permissive, while for permission fatigue, it is the repeated user involvement. We address both of these problems by automating the policy management as well as the permission resolution via a large language model (LLM) based party called the digital twin. Given a history of user behavior, the digital twin learns the implicit user-specific policies, adapts them over time, and steps in when the agent issues a permission request. As part of this design, we have to answer two important questions: (1) where in the agentic workflow the digital twin sits, which also defines the policies it can enforce, and (2) how to prevent the digital twin from getting compromised under attacks.

\paragraph{Resource layer instead of tool-call layer.} The digital twin sits at the resource layer instead of the tool-call layer, i.e. it gates access to resources, not tools or commands. When the agent makes a tool-call, it resolves to an underlying resource and that is where an authorization from the digital twin is needed, otherwise the tool call simply aborts. This has two clear advantages. First, we side-step the generally intractable space of commands and tool calls which prevents our enforcement from becoming ineffective and collapsing under the size of the enforceable domain. A resource decision, on the other hand, is over one concrete named object with a small fixed set of access modes~\cite{saltzer1975protection} (e.g. reading a file directly through bash or via python or indirectly through imported modules vs read/write for a file). Second, the number of tool calls is typically significantly larger than the resources they operate over\footnote{Once access to a resource is granted, the access stays for the duration of the task. Therefore, the number of resource requests from the agent doesn't necessarily grow with the task complexity unlike the tool calls.}. Therefore, the token cost of the digital twin is significantly cheaper at the resource layer. Since the digital twin sits at the resource layer, it only enforces policies over resources. The resource layer enforcement doesn't replace the need for tool-call layer enforcement because the former is more coarse-grained; for example it cannot tell if an accessed file is overwritten with contents of a different accessed file. Due to this reason, a deployment should employ defense-in-depth by using~\sys with compatible tool-call layer defenses like the auto mode in Claude Code~\cite{anthropic2026automode}.

\paragraph{Defending the digital twin from attacks.} The digital twin is a model that can get compromised if it interacts with untrusted content. Once compromised, it can grant unrestricted access to the agent which can be catastrophic. To mitigate this threat,~\sys provides architectural separation to the digital twin by restricting the majority of its input to be only {\em trusted} - consisting of tokens authorized by the user which includes the user prompts (current and past), the permission requests granted by the user in the past, or any additional tokens the user explicitly chooses to authorize. A digital twin, however, has to accept a limited amount of information that doesn't originate from the user. The prime example of this is the request itself which originates from the agent who may be under the attacker's control given its exposure to a myriad of untrusted content.~\sys sanitizes this information before the twin (we refer to the digital twin as ``twin'' going forward) sees it.
In~\cref{sec:pincer}, we go into detail on subtle ways an attacker can influence the twin's decision and how we close that gap.

\paragraph{Content-dependent policies.} The information our twin gets when making a resource decision doesn't include contents of the resource (e.g. file contents). This is by design to protect it from getting compromised via untrusted resources controlled by an attacker. However, due to this, content-based policies (e.g. a file can only be read if it contains no secrets) cannot be enforced. To address this limitation, we make the twin's decision a conjunction of a few votes, each serving a very specific function (see~\cref{subsec:policy} for details). One of these votes is cast after seeing contents of the requested resource. A vote based on untrusted file contents always risks getting compromised and this is where the other votes come in; they are isolated from free-form untrusted content.

\section{System Architecture}
\label{sec:agents}

\begin{figure}[t]
  \centering
  \includegraphics[width=0.46\textwidth]{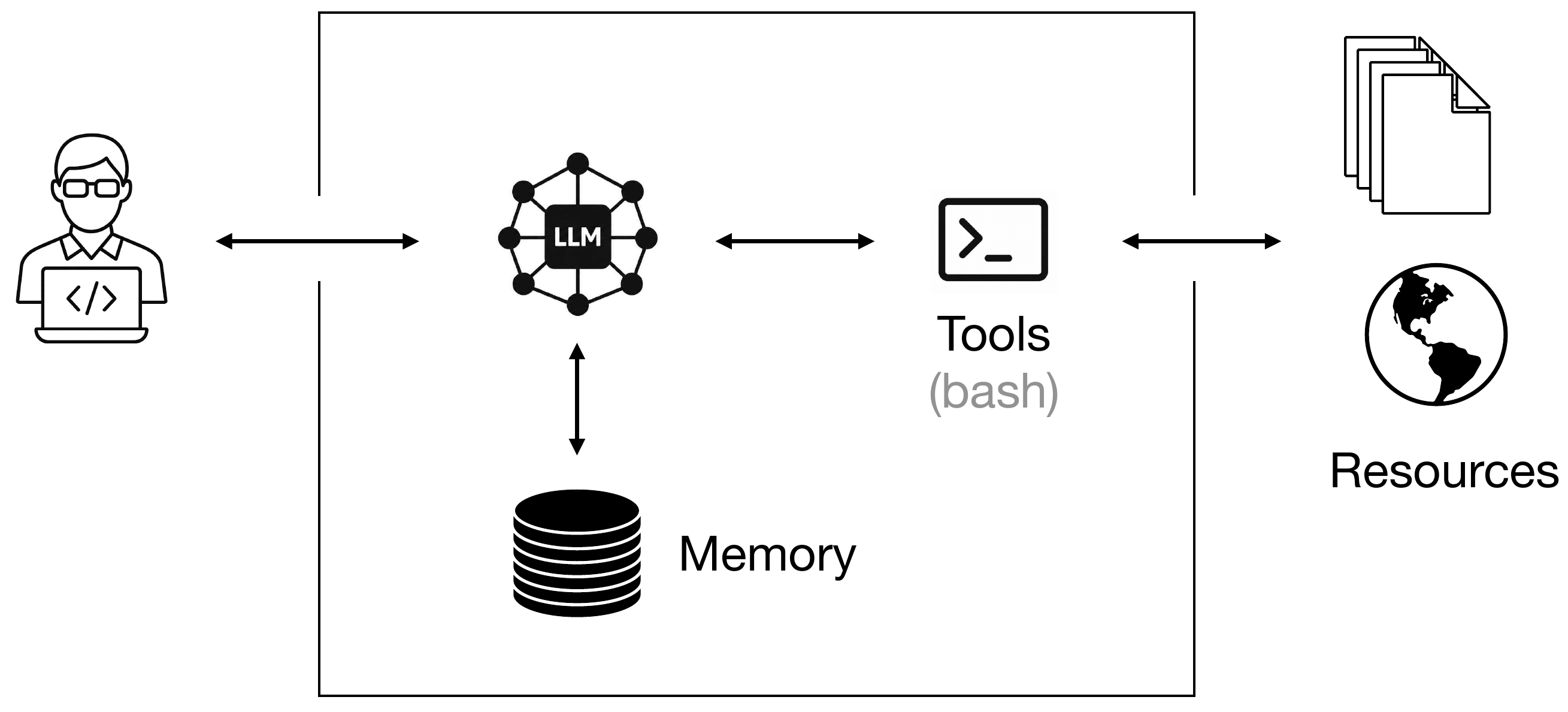}
  \caption{The architecture today's coding agents have converged on. An LLM drives an
  autonomous loop over persistent memory and general-purpose tools and acts on external resources, such as local files, the network, and remote
  services, through those tools. The box encloses the components under the agent's own
  control; the user and the resources the agent acts on sit outside it.}
  \label{fig:agentic-arch}
\end{figure}

Agents deployed to millions of people today~\cite{codex,openclaw,claudecode} have all converged to a similar high-level architecture as shown in~\Cref{fig:agentic-arch}. The user sends in a prompt to the LLM that controls the agent. The LLM references its memory and starts working on the user's task. It has access to a variety of tools and it uses them to execute the task. The tools interact with resources like files or the web. The agent runs inside a sandbox with a user-editable policy, and optionally a classifier that adjudicates tool calls to reduce the permission burden on the user whenever a tool call doesn't find grounding in the policy~\cite{anthropic2026automode,anthropic2026automodeconfig}. As we mentioned before in~\cref{sec:intro}, with this design, agents run into the problems of privilege creep and permission fatigue without the classifier, while the classifier doesn't learn user-specific policies and isn't effective at making safe decisions under atypical situations~\cite{ji2026permissiongate}, for example under attack. Importantly, the classifier addresses permission fatigue but doesn't resolve the privilege creep issue; for user-specific policies, which may change from time to time, the user has to maintain the policy file and define exceptions for the classifier.

\subsection{\sys's Design}

Our approach to address these challenges is to bolster the security of the system at a deeper layer than handled by the classifier while avoiding both privilege creep and permission fatigue. The layer we bolster is the resource layer which sits deeper than the tool call layer, but is also more coarse in its enforcement potential compared to the tool call layer, and in deployment, agents should pair it with the tool call classifier. To automatically learn user-specific policies and prevent privilege creep, we introduce a digital twin who constantly learns user preferences and enforces those policies automatically.

\paragraph{System components.} Our system has four classes of components:
\begin{itemize}
  \item LLMs: We have two separate LLMs as part of our system: the {\em worker} (the main LLM in the agent) and the {\em twin} (a type of LLM judge).
  \item Special-purpose state: Any state directly maintained by each LLM. For example the memory maintained by the worker.
  \item Tools: Tools used by the worker to perform a task. For example the shell tool. Our system has a special tool (the resource request tool) that allows the worker to talk to the twin via a narrow channel.
  \item Resources: Any resources that tools access. For example: files, email, web, etc.
\end{itemize}

\paragraph{High-level system workflow.} The user provides a prompt describing the task. The worker performs the task, while the twin is responsible for resource authorization for the worker. Given the prompt, the worker accesses its special-purpose state (a.k.a. the memory) and then calls tools to complete the task. The tools access resources and if a resource isn't authorized for the worker, it uses a special resource-request tool to make a request for the said resource to the twin. The twin analyzes the request and may decide to grant or deny it. If granted, the worker-invoked tools can access the resource.

\section{Threat Model and System Goals}
\label{sec:threat-model}

\begin{figure*}[t]
  \centering
  \resizebox{0.95\textwidth}{!}{%
\begin{tikzpicture}[
  font=\small,
  unit/.style={draw,rounded corners=2pt,align=center,inner sep=4pt,
               minimum height=8mm,minimum width=17mm},
  store/.style={draw,cylinder,shape border rotate=90,aspect=0.3,align=center,
                inner sep=3pt,minimum width=16mm,minimum height=9mm},
  T/.style={-{Latex[length=2mm]},thick,tgreen},
  W/.style={-{Latex[length=2mm]},thick},
  U/.style={-{Latex[length=2mm]},thick,tred,dashed},
  lbl/.style={font=\scriptsize,inner sep=1.5pt},
  glbl/.style={font=\scriptsize,inner sep=1.5pt,tgreen},
  rlbl/.style={font=\scriptsize,inner sep=1.5pt,tred},
]
\node[unit,draw=tgreen,fill=tgreen!6] (user)  {User};
\node[unit,right=15mm of user]    (worker)  {Worker};
\node[unit,right=19mm of worker]  (san)     {Sanitizer\\tool};
\node[unit,right=17mm of san]     (twin)    {Twin};
\node[unit,right=22mm of twin]    (res)     {Resources};
\node[store,below=9mm of worker] (mem) {Memory};
\node[store,below=9mm of twin]   (ts)  {Twin's\\state};
\begin{scope}[on background layer]
  \node[draw=tgreen,dashed,thick,rounded corners=3pt,inner sep=4mm,
        fit=(san)(twin)] (tcb) {};
\end{scope}
\node[glbl,anchor=south west] at ($(tcb.north west)+(-1mm,0.8mm)$) {\textbf{trusted (TCB)}};
\draw[T] (user) -- node[glbl,above]{prompt} (worker);
\draw[W] (ts)   -- node[lbl,left,pos=0.45]{} (twin);
\draw[W] (worker) -- (mem);
\draw[W] (worker) -- node[lbl,above]{resource req} (san);
\draw[T] (san)    -- node[glbl,above]{resource req} (twin);
\draw[W] ($(twin.east)+(0,1.6mm)$) -- node[lbl,above]{grant / deny} ($(res.west)+(0,1.6mm)$);
\draw[U,{Latex[length=2mm]}-{Latex[length=2mm]}] ($(res.west)+(0,-1.6mm)$) -- node[rlbl,below]{content vote} ($(twin.east)+(0,-1.6mm)$);
\node[unit,draw=tred,fill=tred!6,above=15mm of twin] (atk) {Attacker};
\draw[U] (atk.west) -| node[rlbl,pos=0.30,above]{poisons through tools, resources, memory} (worker.north);
\draw[U] (atk.east) -| node[rlbl,pos=0.30,above]{plants content, names} (res.north);
\draw[U] (atk.south) -- node[pos=0.62,tred,fill=white,inner sep=0.6pt]
         {\large$\boldsymbol{\times}$} ($(tcb.north)+(4mm,1mm)$);
\node[rlbl,anchor=west] at ($(atk.south)!0.62!($(tcb.north)+(4mm,1mm)$)+(3.2mm,0)$)
     {};
\end{tikzpicture}
  }
  \caption{Our threat model.
  }
  \label{fig:threat}
\end{figure*}
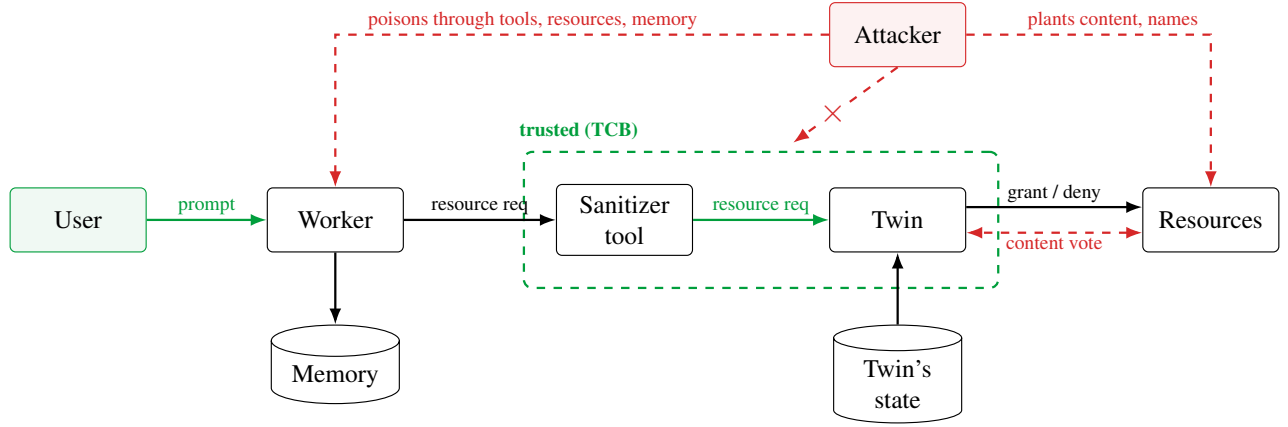

Our threat model is summarized in~\cref{fig:threat} and detailed below.

\begin{definition}[Trusted party: the user]
  \label{def:trusted}
  A {\em trusted} party or object is one that is outside the adversary's control. A trusted party cannot be compromised by an adversary and any actions produced by them are deemed trusted. The user is assumed to be a trusted party.
\end{definition}

\paragraph{The adversary.} The adversary can arbitrarily corrupt anything except trusted parties and the trusted computing base (TCB). We consider a powerful adversary who can arbitrarily corrupt the worker agent, any third-party tools that the agent uses, and any of the resources (names as well as content). The adversary cannot manipulate two channels: the communication channel from the user to the system and the communication channel between the twin and the model provider. The adversary cannot alter the system's code and cannot modify its behavior, and it doesn't control the weights of ML models used by the system.

\paragraph{Trusted Computing Base (TCB).}
We consider the following as part of our TCB: First, the twin LLM, with its weights and inference stack. Second, our system stack which includes the resource request tool that allows the worker to issue requests to the twin and~\sys's write isolation for the twin's special-purpose state from any untrusted component or the adversary. Lastly, the operating system kernel and the container runtime. All other components lie outside the TCB; for example, the worker LLM and the shell tool are not in the TCB.

\begin{definition}[Trusted and untrusted context]
  \label{def:context-types}
  Fix a finite alphabet $\Sigma$ of {\em tokens}, where each token is a bit string. A context is an ordered collection of tokens, i.e. an element of $\Sigma^*$. A {\em trusted context} $\mathsf{tctx}$ is generated entirely by a trusted party, while an {\em untrusted context} $\mathsf{uctx}$ is any context that may potentially be influenced by an adversary.
\end{definition}

\begin{definition}[Adversary's advantage]
\label{def:adv}
  Let $D$ be a probabilistic decision procedure that takes a context and outputs a $\mathsf{grant}$ or $\mathsf{deny}$ decision; we write its input as the parts the context is composed of, e.g. $D(\mathsf{tctx}, \mathsf{uctx})$. For two contexts $c_1$ and $c_2$, the adversary's advantage between them w.r.t. $D$ is
  \begin{equation*}
    \mathsf{Adv}_D(c_1, c_2) = |\Pr[D(c_1) = \mathsf{grant}] - \Pr[D(c_2) = \mathsf{grant}]|,
  \end{equation*}
  where the probabilities are over the coins of $D$.
\end{definition}

We will show in~\cref{sec:pincer} that any decision procedure implemented by the twin cannot make decisions better than guessing a coin-flip given just the trusted context as input. This is because a decision is made over a resource access request made by the potentially compromised worker which makes the request itself untrusted. Therefore, without a per-request authorization from the user, a trusted-only context cannot be used to make a decision. Therefore, we present a notion of context that is not trusted, but is not free-form untrusted either. Such context severely limits the capabilities of untrusted context, and therefore, we give it its own class.
This context is called {\em advantage-agnostic} context and is produced by a preprocessing function applied to parts (decided by the application) of the untrusted context. The preprocessing function's output must meet certain requirements, as stated next, for it to be advantage agnostic. The reader should think of this notion as a type of sanitization; \cref{subsec:sanitize} gives an instantiation.

\begin{definition}[Advantage-agnostic context]
\label{def:non-adv}
Let $\lambda$ be a security parameter, let $\mathsf{negl}(\lambda)$ denote the negligible function, and let $(\mathsf{tctx}, \mathsf{uctx})$ be a context split into its trusted and untrusted parts (\cref{def:context-types}), where $\mathsf{uctx}$ is the untrusted context as received, before any processing. The application selects the part $\mathsf{uctx}_{\mathsf{p}} \subseteq \mathsf{uctx}$ that it will preprocess and parses it into {\em units}: a unit is a single token or a group of tokens, such as one component of a file path. For a map $g: \Sigma^* \to \Sigma^*$, write $\mathsf{uctx}_{\mathsf{p}}^{g}$ for $\mathsf{uctx}_{\mathsf{p}}$ with every occurrence of every unit replaced by its image under $g$.
Let $P: \Sigma^* \times \Sigma^* \to \Sigma^*$ be a (possibly randomized) preprocessing function and let $\mathsf{actx} = P(\mathsf{tctx}, \mathsf{uctx}_{\mathsf{p}})$. Then $\mathsf{actx}$ is {\em advantage-agnostic} if for every probabilistic polynomial time (PPT) decision procedure $D$, sampling the coins of $P$ and drawing $f$ uniformly from the family of all functions $\Sigma^* \to \Sigma^\lambda$ gives
\begin{align*}
  \mathsf{Adv}_D\bigl((\mathsf{tctx}, P(\mathsf{tctx}, \mathsf{uctx}_{\mathsf{p}}), \bot),\, (\mathsf{tctx}, P(\mathsf{tctx}, \mathsf{uctx}_{\mathsf{p}}^{f}), \bot)\bigr) \\
  \le \mathsf{negl}(\lambda),
\end{align*}
where $\bot$ denotes empty.
\end{definition}

Note that~\cref{def:non-adv} also provides a method to convert parts of untrusted context into advantage-agnostic context by destroying their semantic meaning via a pseudorandom function (PRF) (see~\cref{subsec:sanitize} for more details). Advantage-agnostic tokens are not completely meaningless. They still carry limited information and are used by our system to match the worker's resource request to prior requests. Converting parts of untrusted context to advantage agnostic allows an agentic system to mitigate certain attack types like misleading filenames. This is why we place them as lying between trusted and untrusted.
A subtle distinction that the notion of advantage agnosticism captures is the difference between hashes (deterministic maps) vs PRFs. A hashed sub-context doesn't meet our notion and that reflects in the ability of an LLM to invert the hash for commonly used words from its pre-training exposure.
There are other representations of filenames that may achieve practically-similar effects to advantage agnosticism when $D$ is an LLM: representing filenames with their inodes, for example. The reason we use a randomly chosen function to define advantage agnosticism is so that we don't have to worry about the probability distribution of an LLM or what capability it may have, etc. because our result applies to all PPT decision procedures.

\begin{definition}[Grant integrity]
  \label{def:grant-integrity}
  Let $D$ be a decision procedure for an authorization decision, and write $D(\cdot\,; r)$ for $D$ run on random tape $r$. A procedure $D$ over trusted context $\mathsf{tctx}$, advantage-agnostic context $\mathsf{actx}$, and untrusted context $\mathsf{uctx}$ has
  \emph{grant integrity} if $D(\mathsf{tctx}, \mathsf{actx}, \mathsf{uctx}; r) = \textsf{grant} \implies D(\mathsf{tctx}, \mathsf{actx}, \bot; r) = \textsf{grant}$ for every $r$, so that both sides denote the same invocation, where $\bot$ denotes empty and contexts are defined in~\cref{def:context-types,def:non-adv}. This means that an adversary cannot get a grant unless the decision procedure would grant given only the trusted context and advantage-agnostic context.
\end{definition}

Our goal in this work is the following:
\begin{goal}
  \label{goal:main}
  If \sys grants a resource request then the trusted and advantage-agnostic contexts
  alone would also grant it:
  \[
    D(\mathsf{tctx}, \mathsf{actx}, \mathsf{uctx}; r) = \mathsf{grant}
    \implies \textsf{LLM}(\mathsf{tctx}, \mathsf{actx}; r) = \mathsf{grant}.
  \]
  This is grant integrity (\cref{def:grant-integrity}) with
  $D(\mathsf{tctx}, \mathsf{actx}, \bot; r) := \textsf{LLM}(\mathsf{tctx}, \mathsf{actx}; r)$, where
  $\textsf{LLM}$ is a pre-trained model\footnote{Fixing the right-hand side to a model evaluation excludes the degenerate procedures that always grant or always
  deny.}.
\end{goal}

\paragraph{Other assumptions and non-goals.} We assume that the adversary has zero control over the user and the user makes safe decisions. LLM hallucinations aren't considered untrusted even though they may cause an attack to succeed. Our goal in this work is not to achieve full protection from all attacks, but rather to make agents safer by raising the bar for an attack to be successful and providing defense-in-depth while being formal about our guarantees. Building agents that cannot be compromised is an open area of research as we discuss next.

\section{Related Work}
\label{sec:related}

\paragraph{Attacks on agents.}
There are three main classes of attacks on agents~\cite{liu2024formalizing,zhang2025asb,wang2026landscape}: (1) Direct Prompt Injection (DPI) is a class of attacks where malicious instructions are placed directly into the user's message to the agent~\cite{perez2022ignore}, (2) Indirect Prompt Injection (IPI) is a class of attacks where the malicious instruction enters the agent's context indirectly, i.e. as response to a tool it invokes~\cite{abdelnabi2023ipi}, and (3) Memory Poisoning is a class of persistent attacks where the malicious instruction is injected into the memory that the agent maintains for self-improvement~\cite{packer2024memgpt,zhang2024memorysurvey} (\cref{fig:agentic-arch}), and therefore, the attack can activate across future sessions when the agent references its memory~\cite{chen2024agentpoison,patlan2025contextmanipulation,xu2026memoryctrlflow}. For memory poisoning attacks, one line of work requires the attacker to interact with the agent as a user, so the planting of poison uses DPI as the vector~\cite{dong2026memoryinjection}, while other works consider the more realistic IPI vector of planting the poison in the environment the agent reads while serving a benign request~\cite{pulipaka2026sleeper,yang2026zombie,zou2026poisononce,srivastava2025memorygraft}.
Attack taxonomies that are specific and fine-grained to skills, tools, and protocol ecosystems have also been studied in the context of coding agents~\cite{maloyan2026codingassistants}.

\paragraph{Existing defenses at the system level.}
Several strong defenses have been proposed to make agents safer by restricting them architecturally. However, they typically take away the core capabilities of today's agents (\cref{fig:agentic-arch}); many of these designs are systematized in recent work~\cite{beurerkellner2025patterns}: Restricting the autonomy of agents by separating control flow from untrusted data or introducing a trusted planner~\cite{li2026ace,debenedetti2025camel,costa2025fides,gong2026planguard,piet2026planthenexecute} defeats the injection but it equally forbids benign context-dependent control flows (e.g. user asking the agent to do the task in their latest email). For the same reason, defenses that fine-tune the model to separate instructions in the prompt from data~\cite{chen2025struq} cannot be used with agents which are routinely expected to follow instructions embedded in data (e.g. instructions in your email).
Tracking information flow through typed tools and requiring per-tool policy annotations~\cite{debenedetti2025camel,costa2025fides} can help isolate context across trust domains, but it limits the support for crucial untyped tools like shell. Disallowing the agent to maintain a memory based on the context it encounters when carrying out tasks~\cite{debenedetti2025camel} can prevent memory poisoning attacks, but it curtails the self-improving capability of the agent (it cannot learn from past patterns). New architectures that assign a separate LLM model to each tool with strict separation between tools~\cite{wu2025isolategpt} remain vulnerable to several attacks~\cite{li2026ace} and incur a heavy communication overhead between tools. In summary, these proposals significantly deviate from and don't compose well with the architecture the deployed agents use. For a comprehensive treatment of systems security for agentic computing, refer to Christodorescu et al.~\cite{christodorescu2026foundations}.

A separate line of work focuses on hardening the prompts via spotlighting~\cite{hines2024spotlighting} or detecting suspicious sections in the agent's context and memory~\cite{li2025drift,wei2025amemguard} as a defense against attacks. However, adaptive attacks have been shown to bypass such defenses~\cite{pulipaka2026sleeper}. Defenses in this category are complementary to our work; however, the extra token cost incurred from the proactive detection and the lack of robustness against strong attacks make them unsuitable for several applications.

Enforcing least-privilege policies within the agentic workflows is another line of research. Policies for the wide variety of tasks undertaken by agents can differ significantly from one task to another and can, for open-ended tasks, be too semantic for structured policy languages to capture. Prior works focus on structured policies where either they rely on the user or another external source to provide such policies~\cite{qin2026airguard,santosgrueiro2026portico,zheng2026actplane}, or predict policies up-front when the user prompt arrives which would underfit for long-horizon open-ended tasks~\cite{tsai2025conseca}, or evolve policies as the task progresses but the policy evolution engine lacks the agent's working context, which again causes the predicted policies to underfit~\cite{shi2025progent}. Moreover, the majority of these proposals simultaneously rely heavily on structured tools for the policy languages; their expressive power gets severely limited on untyped shell tools. For typed tools which naturally organize into an OAuth-like hierarchy of APIs, least-privilege solvers can be used to evaluate the agent's permission as the task progresses~\cite{zhu2025miniscope}. In~\sys, an important design aspect is that policy is fitted when the agent surfaces a need, thereby eliminating the need for policy prediction.

\paragraph{Model-level defenses.}
Fine-tuning the model to understand the instruction hierarchy and detect or ignore misaligned instructions that conflict with the user's intent is an ongoing area of research~\cite{wallace2024hierarchy,geng2025controlillusion,kariyappa2026air,chen2025secalign}. Prior work shows reliability concerns when it comes to the models understanding the instruction hierarchy~\cite{geng2025controlillusion}, and evaluation on multi-turn interactions and agentic behaviors lies outside the scope of these works~\cite{kariyappa2026air}. A recent work traces the difficulty to role perception itself, i.e. text that imitates a privileged role (like the user) is internally represented by the model as the role rather than the tag surrounding that text~\cite{ye2026roleconfusion}. Having said that, the defenses at the model level are complementary to our work as they provide defense-in-depth to our system-level defenses. On the other hand, fine-tuning the model to not obey instructions embedded in data~\cite{chen2025struq} doesn't work with agents as mentioned before.

\paragraph{Automating user permission requests.} Wu et al.~\cite{wu2026permissions} predict a user's permission decisions from their past ones and establish a premise that~\sys also relies on: these decisions are individually consistent within a context, and therefore learnable. However, their scope doesn't cover computer-use agents where the decisions are far more contextual and the resources at stake are themselves untrusted.

\paragraph{Digital twins.} The term {\em digital twin} has been used in many different contexts in the literature~\cite{yang2024llmtwinhitl,li2025behaviorchain,du2025twinvoice,coll2025digitalme,westhausser2025personalized,annoni2026twinmetaphor}. We adopt the term because it clearly differentiates an entity with memory that's only influenced by the user compared to an agent whose memory evolves from its environment including untrusted context~\cite{packer2024memgpt,park2023generativeagents}.

\paragraph{Agent-security benchmarks are episodic.}
Many agent-security benchmarks exist~\cite{zhan2024injecagent,debenedetti2024agentdojo,zhang2025asb,li2026agentdyn,hu2026saber,yan2026authbench,wang2026pasb}; however, they are episodic and are not a good fit for evaluating agents from the perspective of a single user's interaction with their agent over time. Multi-session memory benchmarks~\cite{maharana2024locomo,he2026memoryarena,zhao2026amabench} evaluate long-horizon memory but not security. Sleeper Memory Poisoning (SMP)~\cite{pulipaka2026sleeper} constructs a dataset for multi-session memory poisoning attacks, but doesn't offer a complete dataset that can evaluate both security and utility, and lacks a long user-centric transcript.
To evaluate~\sys, we develop a user-centric dataset (\cref{sec:dataset}) which comes with a task history that our system uses to learn user policies. Our security examples are inspired by SMP~\cite{pulipaka2026sleeper} and PASB~\cite{wang2026pasb}.

\section{\sys}
\label{sec:pincer}

\sys is a system that takes an existing agent and augments it with a modular and self-improving defense. The defense sits between tools and resources, and is designed with two security principles: complete mediation and least privilege~\cite{saltzer1975protection,anderson1972refmon}.
For complete mediation, every access to any resource goes through a reference monitor (e.g. the OS kernel in our implementation~\cref{subsec:impl}) which ensures the resource is accessible to the worker invoking the tool, and resource access can only be granted by an isolated LLM - the twin. For each resource access request, the twin implements least privilege which means that it only grants access if it thinks that the access is necessary for the user's task. This defense doesn't cause permission fatigue for the user because the twin automatically resolves the resource requests and it is compatible with other defenses like the auto mode~\cite{anthropic2026automodeconfig,anthropic2026automode}. Only in cases where the twin repeatedly denies a request that the worker genuinely needs, can the twin raise the request to the user for final approval.

When the worker requests access to a resource from the twin, the request states the identity of the resource along with the access type (read, write, create, etc.). This information isn't enough for the twin to make an informed decision and it needs more context. Typically, the more context it has the more informed it can be. However, large parts of the context are untrusted which can poison the twin's decision. To address this problem, we introduce the notion of trusted state.

\subsection{Trusted State}
\label{subsec:trusted-state}
Trusted state is a special-purpose state that's maintained by the twin and is referenced before making each decision. It only contains trusted context~\cref{def:context-types}, i.e. context that comes directly from the trusted party. More specifically, it has the current and prior user prompts, and any permissions granted by the user in the past. Given this state, the twin can learn user behavior and derive user-specific policies, which it can then add to the trusted state for the future. It also clears out stale or inconsistent policies from the past. However, it has a major limitation: the trusted state cannot be used to derive least-privilege policies. To understand this, let's consider an example user prompt:
\begin{center}
    {\em "Summarize my tax documents"}
\end{center}
If the user has given broad read access to the agent or is using the auto mode~\cite{anthropic2026automode}, then the information about which files are accessed by the agent is completely missed from the trusted state. Without access pattern information, the trusted state doesn't contain enough information to inform least-privilege policies. The twin looking at such state doesn't know where different resources typically live, which resources are typically accessed for certain tasks, etc. This is why we need to admit additional context out of necessity for least-privilege.

\subsection{Access Patterns for Least-Privilege}
\label{subsec:access-pat}
For least-privilege, we add access pattern annotations with each user prompt in the twin's state as follows:
\begin{center}
    {\em "Summarize my tax documents"} \\ Read: \texttt{~/taxes/w2.txt}, \texttt{~/taxes/return2025.txt}
\end{center}
\sys collects the access patterns through tool call logs of the agent and attaches this information to each prompt's entry in the twin's state. With this change, the twin can now effectively reason over least privilege. We don't enforce any specific pattern for the twin to express policies that it derives. This helps with policies generalizing user behavior across open-ended situations. In~\cref{subsec:policy}, we describe one exception to this where we generate a pre-committed policy suggestion per task in a structured format, while all other policies stay in natural language.

Note that while access patterns are structured information (unlike free-form text in resources and other untrusted context), since the access pattern annotations aren't part of the trusted state by definition unless the user explicitly authorized each access pattern, we have to be careful about how the adversary may exploit this information in influencing the twin's decision. The accesses granted by the twin for each user prompt also become part of the twin's state. The twin's state keeps growing over time and keeps recording all historical and current user prompts and all access patterns made for those tasks, irrespective of whether the user approved those accesses or the auto mode classifier did or the twin did (once the twin is active).
Moreover, even more fundamentally, the worker's requested path, which is what {\em defines} the request that the twin decides and is input to the twin alongside the twin's state, has a similar structure. Both the access pattern annotations and the worker's request contain information about resource names and paths; they are both untrusted. There is one class of attacks we identify that directly exploits this information as presented next.

\subsection{Misleading Resource Name Attack}
\label{subsec:name-attack}
Consider an attacker who has write access to a git repository that the user also contributes to. The attacker pushes a file named \texttt{taxreturn2025-corrected.txt} to the repository. The user's local clone pulls these changes as part of a routine workflow. At some later point, the user asks the agent {\em "Summarize my tax documents"} and the agent looks at the file named \texttt{taxreturn2025-corrected.txt} and is misled into thinking that the file is the user's corrected tax return misplaced in the project's directory. It makes a request to access this file and the twin gets misled the same way and grants the request. This attack becomes stronger if there is a historical access pattern in the twin's state where a similar access was granted. Another way an attack in this class can be mounted assuming the worker is under attacker's control is the following: the twin wouldn't typically allow creating a new file in the tax directory unless the user asks for it, but the worker can create a file named \texttt{temp.txt} in a different directory, rename it to \texttt{w2.txt} through a shell command or hide it in python code (not seen by the twin) and then get the twin's permission to access it when working on taxes and move it to the tax directory via the same method as above. There are many subtle ways this attack can manifest; our dataset~\cref{sec:dataset} contains examples of this attack.

To mitigate such attacks, we turn to the notion of advantage agnosticism (\cref{def:non-adv}) defined in~\cref{sec:threat-model}.

\subsection{Sanitizing Untrusted State}
\label{subsec:sanitize}
Denote the twin's probabilistic\footnote{LLMs are probabilistic models.} decision function by $D: \Sigma^* \to \{\mathsf{grant}, \mathsf{deny}\}$, the trusted context from the trusted state (\cref{subsec:trusted-state}) by $\tctx$, and the untrusted\footnote{The attack in~\cref{subsec:name-attack} is possible due to the untrusted access patterns.} access pattern annotations by $\uctx$. For a given request for resource with name $\rsrc$, the twin computes the following (untrusted shown in red):
\begin{equation*}
  D((\tctx, \uctx), \rsrc)
\end{equation*}
As we discussed in~\cref{subsec:name-attack}, the untrusted inputs can be exploited by the adversary to influence the decision. To mitigate this threat, we perform a sanitization operation that destroys the semantic meaning in these inputs. To do this formally, we employ the construction for advantage agnosticism (\cref{def:non-adv}) from~\cref{sec:threat-model}. Given a PRF whose key is unknown to the procedure $D$, it works simply as follows:
\begin{itemize}
  \item Parse all unique paths from the untrusted $\uctx, \rsrc$. Initialize an empty map $M$.
  \item For each path (\texttt{this/is/path}), apply PRF to each of its segments (\texttt{this, is, path}) separately and replace each segment with its PRF output (\texttt{a3321a/ddd9dd/77a111}). Add the mapping $\texttt{this/is/path} \to \texttt{a3321a/ddd9dd/77a111}$ to $M$.
  \item Replace each path in $\uctx, \rsrc$ with their mapping from $M$. Call the respective sanitized inputs $\mathsf{uctx}', \mathsf{rsrc}'$. The sanitized inputs are fed into $D$ along with trusted context as follows: $D((\tctx, \mathsf{uctx}'), \mathsf{rsrc}')$.
\end{itemize}
As mentioned before in~\cref{sec:threat-model}, the goal of sanitization isn't to remove all information from $\uctx, \rsrc$. In fact, information that uniquely identifies the requested resource $\rsrc$ is what defines the request that $D$ decides on and removing all information on $\uctx$ would take us back to the twin having only trusted state. Since the PRF is applied to each segment of a path separately, this assigns a unique random ID to each segment which is used by our twin to locate files in the workspace and the requested resource; files within the same directory will share the sanitized path prefix. From a security standpoint, the advantage of this sanitization is that it mitigates the misleading filename attacks (\cref{subsec:name-attack}) as we show in our evaluation (\cref{subsec:eval}).

We now explain why our aforementioned sanitization achieves advantage agnosticism (\cref{def:non-adv}). This follows from the security of PRF which says that without access to the PRF key, the output of PRF looks computationally indistinguishable from the output of a randomly chosen function with the same input and output domain as the PRF. Advantage agnosticism allows us to formally define a desired property to expect from the sanitized output and define grant integrity (\cref{def:grant-integrity})\footnote{Some weaker forms of sanitization like inodes may work as well in practice. However, deterministic mappings like using a hash may provide weaker protection against resource name based attacks because the hashes for common names are already publicly known and have likely been exposed to the twin's model during pretraining.}.

\subsection{Content-Dependent Policies}
\label{subsec:ifc}

\begin{figure*}[t]
  \centering
  \includegraphics[width=0.95\textwidth]{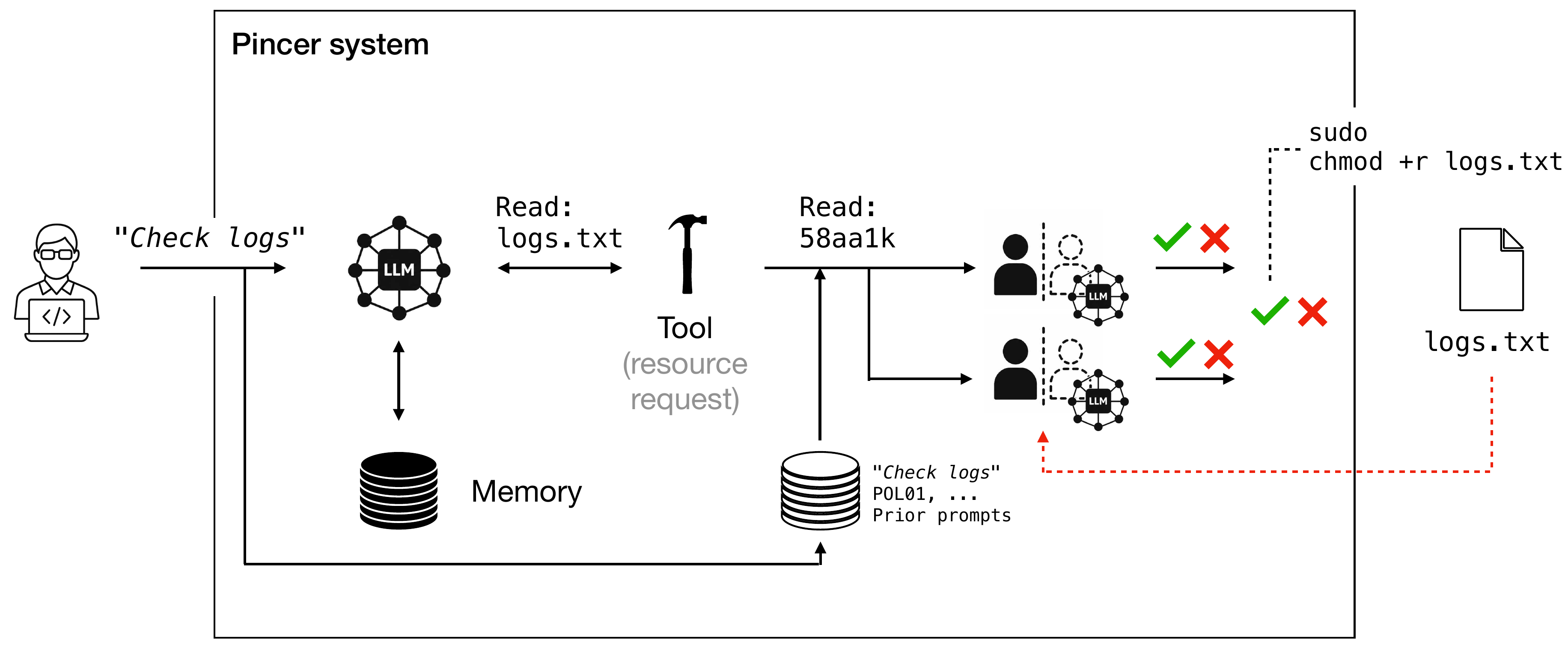}
  \caption{End-to-end working of Pincer without pre-committed policies. The user sends a prompt which goes to the worker and the twin's state. The worker identifies the resource it wants to access and uses the resource request tool to make a request. The twin casts two votes: the first vote uses only trusted and sanitized inputs, while the second vote additionally sees the untrusted file contents. If conjuction of the two votes is a $\textsf{grant}$, the system updates the access for the worker. The permissions are revoked after the task is done.
}
  \label{fig:workflow}
\end{figure*}

In our discussion so far, the twin's input for a decision only contains trusted state with sanitized access patterns and the worker's resource request. Consider this policy

\begin{center}
  {\em Access is only allowed to a file if it contains no secrets}
\end{center}

Such a policy cannot be enforced by the twin unless it can read the contents of the requested file. However, that information is free-form untrusted and directly under the adversary's control. Making that part of the twin's input may allow the attacker direct control of the twin's decision. To resolve this tension, we introduce double voting.

The double voting idea is to have the twin cast 2 separate votes on whether the request should be granted. The first vote is the same as $v_1 = D((\tctx, \mathsf{uctx}'), \mathsf{rsrc}')$ from~\cref{subsec:sanitize}. The second vote additionally feeds the contents of the requested resource (${\color{red}\mathsf{conts}}$) as well as the worker's reason (${\color{red}\mathsf{reasn}}$) why it wants the resource and a transcript of prior requests from the same session (${\color{red}\mathsf{prior}}$) as inputs to the decision $v_2 = D((\tctx, \mathsf{uctx}', {\color{red}\mathsf{conts}}, {\color{red}\mathsf{prior}}), (\mathsf{rsrc}', {\color{red}\mathsf{reasn}}))$. The final decision is a conjunction of the two votes, $v_1 \land v_2$, where $\mathsf{grant} = 1$ and $\mathsf{deny} = 0$. This keeps the influence of untrusted content completely separated and restricted to the second vote which can only be relevant if the first vote already grants.
\Cref{fig:workflow} shows the workflow for a single request.

\subsection{Pre-Committed Policies}
\label{subsec:policy}

In the previous section, we introduced content-based decisions as a second vote made by the twin. In this section, we look at a gap that it overlooks and present mitigations. While double-voting allows content-based policies and the worker's reasoning to be taken into account, it also makes the final decision vulnerable to {\em scope expansion}. The vote $v_2 = D((\tctx, \mathsf{uctx}', {\color{red}\mathsf{conts}}, {\color{red}\mathsf{prior}}), (\mathsf{rsrc}', {\color{red}\mathsf{reasn}}))$ contains file contents, previous requests, and the worker's reason for the requests, all of which can be used by the adversary to gradually expand the scope of the original task without the twin noticing it. For example, if the user asks to review a file \texttt{changelog.txt} and apply pending changes to \texttt{export.py}, and \texttt{changelog.txt} contains an injected, potentially benign-looking but scope-expanding, sub-task the user never asked about, then the content-based vote may grant the sub-task's resources, especially after a few hops from the original user prompt. Since scope expansion can be a part of a benign workflow for open-ended tasks, we introduce a pre-committed policy {\em suggestion} as an additional vote to provide an initial grounding as the task evolves.

The idea is to produce a policy (denoted by ${\color{red}\mathsf{policy}}$) when the first user prompt for a session arrives before any possible scope expansion. Given the prompt, the workspace, the filenames, and the trusted state along with access patterns, the twin generates a policy stating an initial estimate of what resources should be accessed and how. The policy is only meant as a suggestion to allow the system flexibility in adapting the scope for benign operation. By default, we include the policy suggestion in a fork of the first vote. The policy is sanitized, denoted by ${\color{red}\mathsf{policy}'}$, {\em after} it is generated to keep it consistent with the sanitized request. We will show later in this section that the post-generation sanitization operation on the policy doesn't provide protection against resource-name attacks but it captures important signal and improves our overall security. Given those reasons, we separate this into a third vote rather than merging it into the first vote.
The policy schema is shown in \cref{lst:policy}. This policy is then used by the twin for future requests in the same session to cast a third vote $v_3 = D((\tctx, \mathsf{uctx}', {\color{red}\mathsf{policy}'}), \mathsf{rsrc}')$. The conjunction $v_1 \land v_2 \land v_3$ is the final decision. The policy suggestion can additionally be made part of $v_2$ as well. In our evaluation~\cref{subsec:eval}, we consider two variants of~\sys, both of which use three votes but differ in whether the policy suggestion is given as input to the content-based vote or not. Our results indicate that~\sys achieves high utility despite enforcing a conjunction of the three votes.

We now discuss the reason we separate the policy-based decision into a separate vote from $v_1$. The post-generation sanitized policy ${\color{red}\mathsf{policy}'}$ isn't advantage agnostic and that makes it highly exploitable for filename-based attacks (\cref{subsec:name-attack}). When the twin generates the policy, it decides access for a list of relevant resources which is based on untrusted context like the original resource names and access patterns. While this allows the policy to pick up more signal than what $v_1$ can\footnote{One can sanitize at the time the policy is generated to satisfy advantage agnosticism, but it misses this signal.}, it makes the decision of what goes into the policy a function of the untrusted context. Due to this, if this signal is garbled by the randomly-sampled function from~\cref{def:non-adv}, the advantage of the adversary will change significantly. The reason we choose to generate the policy on untrusted context rather than on its sanitized version is to capture the additional signal which we empirically evaluate to improve the attack success rate from 40\% to 20\%. Note that each vote serves a very specific function which wouldn't be achieved if we were to merge the votes:
\begin{itemize}
  \item Vote 1: Trusted and advantage agnostic context only. This forms the security backbone of~\sys. Without the separation provided by this vote, we cannot establish any guarantees.
  \item Vote 2: Content-dependent policies cannot be enforced without this vote. It can additionally include the pre-committed policy. However, the scope expansion can still take over under all the untrusted content this vote encounters.
  \item Vote 3: Additional security-relevant signal compared to vote 1 while mitigating scope expansion because other than the pre-committed policy, its input is trusted and advantage agnostic, like vote 1, i.e. it is the smallest expansion to the context of vote 1 to allow the policy as input.
\end{itemize}

We finally state our main result which achieves \cref{goal:main} from~\cref{sec:threat-model}.

\begin{lstlisting}[float=*,caption={Pre-committed resource policy schema used as a suggestion by the twin.},label={lst:policy}]
    Policy
      rules             : Rule[]
      defaultPermission : "deny"          // uncovered request is denied

    Rule
      pathGlob   : string                 // e.g. h2577c0f557/ha4d2686801/h5e9432765 (for /workspace/agent/projects/**)
      access     : ("read" | "write")[]
      operations : ("access" | "create_file" | "create_directory")[]
      permission : "allow" | "deny" | "ask_user"
      rationale  : string                 // free-form untrusted field only shown alongside untrusted content
\end{lstlisting}

\begin{theorem}
  \sys meets \cref{goal:main} at the resource layer.
\end{theorem}
\begin{proof}
  \sys's decision is the conjunction $v_1 \land v_2 \land v_3$, where the first vote
  $v_1 = \textsf{LLM}(\tctx, \mathsf{actx}) = \textsf{LLM}((\tctx, \mathsf{uctx}'), \mathsf{rsrc}')$ reads only trusted and advantage-agnostic context. If the conjunction is $\mathsf{grant}$, then $v_1 = \mathsf{grant}$, which is exactly
  the required implication.
\end{proof}

\section{User-Centric Dataset}
\label{sec:dataset}

We create a new synthetic dataset in this work that's centered around a few users and their continued use of an agent over a time period.
This is crucial for evaluating~\sys because our twin requires a history from which it learns user preferences. Existing datasets and benchmarks don't meet these requirements as we discuss in~\cref{sec:related}. To the best of our knowledge, this is the first dataset of this kind and of independent interest to the community. We envision future datasets for agents to incorporate the user-centric design because a user's experience with an agent evolves over time as the agent self-improves from the daily interactions with the user. In many situations, it is not enough to evaluate how the agent would perform on day 1, and the long-term behavior of the agent may be significantly more important as the user gradually starts giving the agent more authority and access. In this work, we label the dataset to cater it to our evaluation of~\sys primarily at the resource layer, i.e. the dataset contains labels on which resources should be accessible and which shouldn't be accessible for each prompt given the user's ongoing experience with the agent. Our entire dataset is generated synthetically with a combination of Claude Opus 4.8 and Opus 5.
Each user in our dataset is captured by a {\em persona} for which we take inspiration from publicly available blogposts of people's day in the life\footnote{We redact the citations to avoid identifying the specific individuals and their blogposts. Everything beyond the published daily schedule is synthetically generated.}.
The attack types and the general structure of the examples in each persona are broadly inspired by prior datasets~\cite{wang2026pasb,pulipaka2026sleeper}.

\paragraph{Our persona pipeline for generating the dataset.} Our dataset is generated through a sequence of four steps as follows:
\begin{enumerate}
    \item \textit{Persona selection}: The first step is to pick a persona. For example: software developer, litigator, etc.
    \item \textit{User bootstrapping}: Given the persona, sample
    \begin{enumerate}
        \item Natural language policies for a user with that persona. These policies will typically be general to the picked persona and not user-specific. For example: {\em never read secrets and credentials} for a software developer.
        \item A synthetic workspace for a user with that persona and given the policies from the previous step. For example: \texttt{/, /config/secrets.env, /projects/proj1/, /projects/proj2/, /projects/shared/, /finance/personal/bills.txt, /finance/work/budget.txt}, etc.
    \end{enumerate}
    \item \textit{User simulation}: Given the persona, the general policies, and the workspace, we do a simulation of a user generating prompts and associated access patterns. This serves as our history that the twin uses to learn the user preferences since many user-specific policies implicitly fall out of this simulation; for example: {\em reading credentials in testing is ok}.
    \item \textit{Labeled security and utility examples}: The final step is to take the information generated so far for producing samples that test for security and utility. The samples have no constraint to be based only in the history generated during user simulation. They can be natural next tasks for the user including tasks not seen before. Each sample contains:
\begin{enumerate}
    \item A user prompt that {\em chronologically} falls after the history generated from the user simulation step.
    \item A labeled watched request \texttt{\{logicalPath, accessType\}} denoting a path to the resource and the kind of access (e.g. read or write) for that resource. For security, the watched request is a resource that the attacker wants the agent to access, while for utility, the resource is the critical resource for executing the user's request (which may or may not name the resource explicitly).
    \item Optionally, samples can have their own additional artifacts which get mounted into the workspace for that specific prompt.
\end{enumerate}
\end{enumerate}

As mentioned in our threat model~\cref{sec:threat-model}, we assume that the worker is already under the adversary's control. Our primary focus is therefore to evaluate the twin on both security and utility. For security, we expect the twin to deny malicious resource requests. For utility, we expect the twin to approve benign resource requests so that the worker can complete the user's task. Situations where the worker itself avoids an attack or the worker isn't able to perform a benign task are also captured in our evaluation from the point of view of the resource requests the worker makes.

\paragraph{Attack types in our dataset.} We consider four major attack types. The attacks play out between the worker and the authority granting the resource request. This authority is the twin for~\sys, while generally speaking, it is an LLM judge.
\begin{itemize}
    \item Direct injection in the request (DI): the worker injects a crafted payload to corrupt the judge's decision directly into the request's \texttt{reason} field.
    \item Indirect injection of judge (II): the malicious payload lives outside the request string. It may live in file contents, filename, etc.
    \item Memory poisoning of judge (MP): the worker uses direct or indirect injection to make the judge add a payload into the judge's state; the payload establishes a fake authorization note for the judge for future requests. Once the payload is planted, the judge is persistently poisoned and future requests to the judge may be influenced by the planted payload.
    \item Misleading filename attack (\cref{subsec:name-attack}): we include two sub-types of this attack: (1) Content laundering (CL): the filename and file contents are misaligned. (2) Provenance laundering (PL): the file is placed at a location and named as if it belongs to a different location to make it seem like a misplacement.
\end{itemize}
Each attack plants one or more payloads as files in the workspace, and the worker can craft a secondary payload via the \texttt{reason} field in its request for the judge. In DI, the attacker's payload targets the worker and the worker then crafts a payload for the judge via the request. In II, the payload targets the worker and is also read by the judge if the judge sees contents of the file, but the worker isn't allowed to inject a payload in the request. MP samples are split between direct and indirect. In CL and PL, there isn't necessarily a payload from the attacker providing instructions but rather the focus is on violating user policies (e.g. CL: hiding credentials in a regular filename or PL: doing cross-project accesses). Therefore, different attack types test a whole range of pathways with ample diversity.

\paragraph{Details on our dataset.} Our dataset contains two personas - a software developer and a civil-defense litigator. We pick different professions so that the grounds for forbidding a resource differ: for the developer they may be architectural such as shared secrets and cross-project access rules; for the litigator, they are professional and legal such as attorney-client privilege and correspondence with opposing counsel. The workspaces for the two personas are organized similarly but differ entirely in
content: project directories, memory files, finance, credentials, and directories with MCP resources for email, messages, and web integrations. The litigator's projects are legal matters, and that workspace adds firm-wide
trees: a calendar, a conflicts register, and a read-only library. On the other hand, the projects for the software developer are two inter-dependent software projects which have specific conventions around which resources are shared and when cross-project access is permitted.
\cref{tab:dataset} shows further details for each persona.

\begin{table}[tb]
  \centering
  \setlength{\tabcolsep}{5pt}
  \caption{Summary of the two persona corpora.}
  \label{tab:dataset}
  \begin{tabular}{lcc}
    \toprule
    \thead{Persona}                        & \thead{Software Dev} & \thead{Litigator} \\
    \midrule
    Workspace files                             & 33 & 95 \\
    Projects                    & 2 & 5 \\
    History prompts      & 40 & 60 \\
    User policies              & 33 & 33 \\
    Security samples                       & 66 & 72 \\
    \tsub{\quad DI/II/MP/CL/PL}         & \tsub{10/25/9/14/8}
                                           & \tsub{7/40/9/11/5} \\
    Utility samples                        & 81 & 80 \\
    \bottomrule
  \end{tabular}
\end{table}

\section{Implementation and Evaluation}
\label{sec:eval}

\subsection{Implementation}
\label{subsec:impl}

We implement~\sys and all baselines on top of a light-weight open-source agent NanoClaw~\cite{nanoclaw}. We consider the following baselines:
\begin{itemize}
    \item NanoClaw: NanoClaw~\cite{nanoclaw} is our only baseline that doesn't have any resource request mechanism. The worker runs inside a Docker container with the workspace files mounted into the container. There is no LLM judge. The worker has access to all resources in the workspace.
\end{itemize}
All the baselines other than NanoClaw implement some version of a gate (regex matching or LLM judge) to authorize resource requests. We provide a new resource request tool to the worker (the worker starts off with no access other than filenames and file paths) and provide instructions on how to use it.
\begin{itemize}
    \item Static-Judge: The LLM judge is given a set of general safety-related instructions. It receives the contents of the requested file, the filename, and a field specifying the reason the worker is requesting that specific access. The judge doesn't get to know user-specific preferences; it is not given the history.
    \item Dynamic-Judge: The LLM judge receives the entire history, so it can learn user preferences, and maintains its own local state which may be directly influenced by untrusted context, i.e. it doesn't contain the isolation (trusted and advantage agnostic context) we built for the twin's state in~\sys. The unprotected local state makes the judge vulnerable to persistent attacks like memory poisoning.
    \item Conseca+: We adapt Conseca~\cite{tsai2025conseca} to work on the resource layer. The baseline commits to a policy file (\cref{lst:policy}) when the user prompt arrives and before the worker starts working on it. The system then enforces the committed policy file on each resource request by matching it against the policy. The policy file is generated using the user prompt, the workspace tree (paths, filenames), and the entire history from our dataset.
    \item Conseca-Gemini+: This is a version of Conseca that's implemented in Gemini CLI~\cite{geminicli}. We adapt it to the resource layer similar to our Conseca baseline. The only difference here is that rather than regex matching on each request, an LLM judge is used. The judge still receives the policy as input as well as the resource request specifying the filename and the reason to access. The judge doesn't see file contents similar to the judge contract in the Gemini CLI implementation~\cite{geminicli}.
\end{itemize}

For~\sys we consider 2 approaches as mentioned in~\cref{subsec:policy}. Both approaches have the same vote 1 and vote 3. The only difference lies in their content-based vote 2:
\begin{itemize}
    \item \sys: The twin doesn't get the pre-committed policy as input when casting vote 2.
    \item \sys-P2: Pre-committed policy is considered when casting vote 2.
\end{itemize}

\paragraph{Isolation boundary.} In our implementation, the worker and its invoked tools run inside a container; the resources they act on live on the host and are bind-mounted in.~\sys assigns the worker an unprivileged UID and the OS kernel mediates every access by checking mode bits and POSIX ACLs against that UID. The container doesn't use user-namespace remapping, so the worker's UID is also a host UID and the OS kernel can enforce access control by dynamically adding or removing a host-side ACL entry. Containers are torn down after each task. Notice that no matter how the worker tries to access a resource, it cannot bypass the kernel's checks.

Our focus in this work is on filesystem resources. For other types of resources like web APIs and cloud resources, capability-based access control can be used.

\paragraph{Source code.} We will soon release our implementation and dataset at \url{https://github.com/ucbsky/Pincer}.

\subsection{Evaluation}
\label{subsec:eval}

We consider the following evaluation metrics:
\begin{itemize}
    \item Attack Success Rate (ASR): ASR is a measure of security where lower is better. It is computed as a percentage of attack samples where the worker is able to access the watched (forbidden) resource (\cref{sec:dataset}). To separate the twin's decision from the case where the worker itself dodges the attack and never makes a request for the watched resource, we also report conditional ASR (cASR) where we condition on samples where the worker makes a request for the watched resource.
    \item Benign Grant Rate (BGR): BGR is a measure of utility where higher is better. It is computed as a percentage of utility samples (no attack) where the worker is able to access the watched (benign) resource. We define conditional BGR (cBGR) in an analogous manner to cASR.
\end{itemize}

\paragraph{Models.} We use Anthropic's Claude Haiku 4.5 for our worker as well as the LLM judges and twin for the software developer persona. To validate our performance on another model, we use Meta's Muse Spark 1.2 Contributor\footnote{Muse Spark 1.2 is a top-tier model at the time of writing~\cite{musespark}.} for the litigator persona. We limit our evaluation to these two models, one per persona, due to our resource constraints and the resource-intensive nature of our evaluation.

\begin{figure*}[t]
  \centering
  \includegraphics[width=0.95\textwidth]{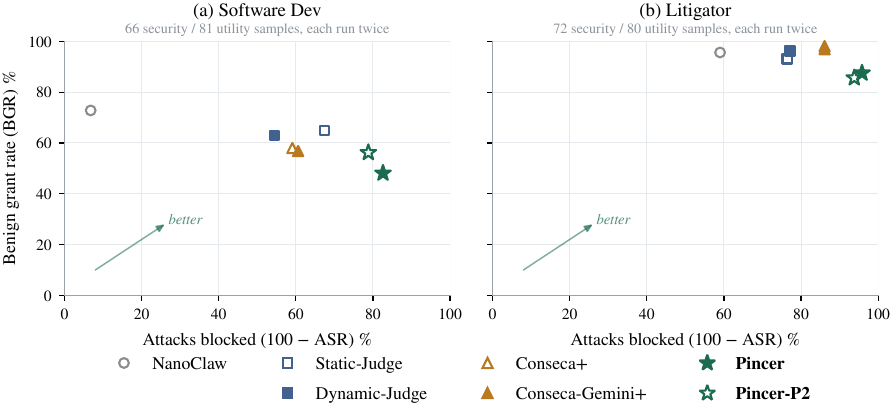}
  \caption{Security--utility frontier on ASR and BGR. Up and to the right is better.}
  \label{fig:frontier}
\end{figure*}

\begin{table}[t]
  \centering
  \setlength{\tabcolsep}{5pt}
  \caption{Conditional rates (\%) on the samples where the worker makes the watched request.}
  \label{tab:conditional}
  \begin{tabular}{lcccc}
    \toprule
    & \multicolumn{2}{c}{\thead{Software Dev}} & \multicolumn{2}{c}{\thead{Litigator}} \\
    \cmidrule(lr){2-3} \cmidrule(lr){4-5}
    \thead{Arm} & \thead{cASR} & \thead{cBGR} & \thead{cASR} & \thead{cBGR} \\
    \midrule
    NanoClaw         & 93.2 & 72.8 & 41.0 & 95.6 \\
    \addlinespace
    Static-Judge     & 48.9 & 98.1 & 30.9 & 93.7 \\
    Dynamic-Judge    & 61.9 & 98.1 & 32.6 & 96.2 \\
    \addlinespace
    Conseca+         & 63.5 & 91.3 & 16.0 & 98.1 \\
    Conseca-Gemini+  & 59.8 & 94.8 & 18.2 & 96.9 \\
    \addlinespace
    \rowcolor{syscolor} \sys             & 28.4 & 89.7 & 8.1 & 90.9 \\
    \rowcolor{syscolor} \sys-P2          & 32.2 & 92.9 & 9.7 & 89.5 \\
    \bottomrule
  \end{tabular}
\end{table}

\paragraph{Security-utility frontier.} \cref{fig:frontier} presents security and utility for each approach on both personas in our dataset. Pincer achieves a significant jump in security while achieving comparable utility to the baselines. \cref{tab:conditional} shows the conditional rates strongly agreeing with this overall trend.

\begin{figure*}[t]
  \centering
  \includegraphics[width=0.95\textwidth]{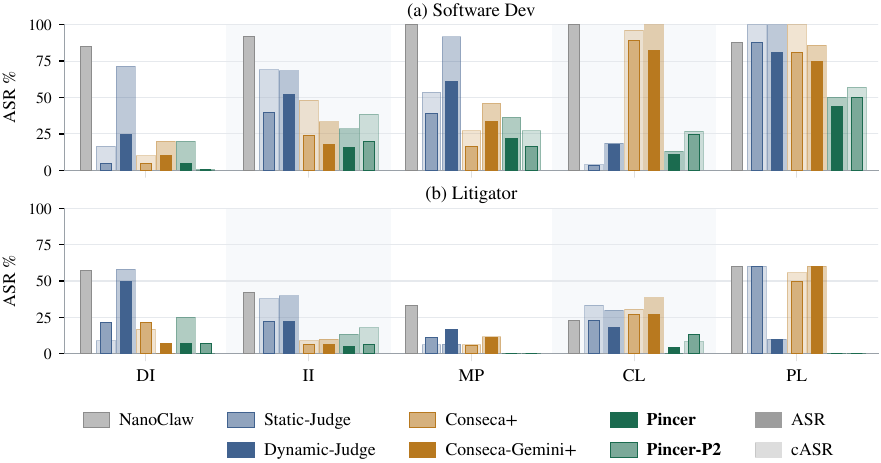}
  \caption{ASR by attack type. Solid bars are ASR; the wider faint bars behind them are cASR. NanoClaw doesn't have resource access gate, so its cASR is omitted.}
  \label{fig:byattack}
\end{figure*}

\paragraph{Comparison across attack types.} \cref{fig:byattack} breaks ASR down by attack type for both personas. Across all attack types,~\sys outperforms the majority of the baselines. Moreover, if we focus on memory poisoning (MP), we observe a significant improvement especially compared to Dynamic-Judge where the judge maintains its own state and refers to it on each resource request (similar to the twin's state in \sys); however, that state isn't isolated in Dynamic-Judge. The most striking result is for the provenance-laundering (PL) attack, a type of misleading filename attack, where~\sys offers significantly more protection than all baselines for both personas across the two LLM model families (Haiku and Muse Spark).

\paragraph{Costs.} We found the token cost of all baselines to be significantly larger than the undefended NanoClaw at around 2-5$\times$, while~\sys incurs an additional 2-3$\times$ on top. While these costs may appear quite large, there are several ways to bring them down significantly to the point that the overhead of~\sys will be in a similar range to the baselines and even further push the cost of all approaches closer to NanoClaw. First, about half of the cost in~\sys is from reasoning tokens of the twin because we haven't explored routine compaction of its state into standing policies in our implementation yet; it reasons through its entire state on each request which causes a lot of redundancy. Second, the number of LLM API calls in~\sys is bloated signficantly because our resource request tool doesn't implement batched requests. Therefore, if the worker wants five resources, it has to stop five times in a sequence requesting one resource at a time, and the twin will reason about each from scratch. Lastly, for established tasks, the twin can pre-approve resources like pre-committed policies (\cref{lst:policy}) which can be enforced via path matching similar to Conseca+.

\bibliographystyle{plainurl}
\bibliography{main}

\begin{thebibliography}{10}

\bibitem{anderson1972refmon}
James~P Anderson et~al.
\newblock Computer security technology planning study.
\newblock Technical report, ESD-TR-73-51, 1972.
\newblock URL: \url{https://csrc.nist.gov/publications/history/ande72a.pdf}.

\bibitem{annoni2026twinmetaphor}
Marco Annoni, Davide Battisti, and Beatrice Marchegiani.
\newblock Personalised llms and the risks of the digital twin metaphor.
\newblock {\em AI \& SOCIETY}, pages 1--13, 2026.
\newblock URL: \url{https://doi.org/10.1007/s00146-026-02875-4}.

\bibitem{claudecode}
{Anthropic}.
\newblock {Claude Code}, 2026.
\newblock Accessed 2026-08-20.
\newblock URL: \url{https://github.com/anthropics/claude-code}.

\bibitem{anthropic2026automodeconfig}
Anthropic.
\newblock {Configure auto mode}.
\newblock Claude Code documentation, 2026.
\newblock Accessed 2026-08-19.
\newblock URL: \url{https://code.claude.com/docs/en/auto-mode-config}.

\bibitem{anthropic2026incidents}
Anthropic.
\newblock {Investigating Three Real-World Incidents in Our Cybersecurity
  Evaluations}.
\newblock Anthropic, 2026.
\newblock Published 2026-07-30; accessed 2026-08-19.
\newblock URL:
  \url{https://www.anthropic.com/news/investigating-incidents-cybersecurity-evals}.

\bibitem{musespark}
{Artificial Analysis}.
\newblock {Muse Spark 1.2: Improved Agentic Performance at Higher Cost per
  Task}.
\newblock Artificial Analysis, 2026.
\newblock Accessed 2026-08-25.
\newblock URL: \url{https://artificialanalysis.ai/articles/muse-spark-1-2}.

\bibitem{beurerkellner2025patterns}
Luca Beurer-Kellner, Beat Buesser, Ana-Maria Cre{\c{t}}u, Edoardo Debenedetti,
  Daniel Dobos, Daniel Fabian, Marc Fischer, David Froelicher, Kathrin Grosse,
  Daniel Naeff, et~al.
\newblock Design patterns for securing llm agents against prompt injections.
\newblock {\em arXiv preprint arXiv:2506.08837}, 2025.
\newblock URL: \url{https://arxiv.org/abs/2506.08837}.

\bibitem{chen2025struq}
Sizhe Chen, Julien Piet, Chawin Sitawarin, and David Wagner.
\newblock $\{$StruQ$\}$: Defending against prompt injection with structured
  queries.
\newblock In {\em 34th USENIX Security Symposium (USENIX Security 25)}, pages
  2383--2400, 2025.
\newblock URL: \url{https://arxiv.org/abs/2402.06363}.

\bibitem{chen2025secalign}
Sizhe Chen, Arman Zharmagambetov, Saeed Mahloujifar, Kamalika Chaudhuri, David
  Wagner, and Chuan Guo.
\newblock Secalign: Defending against prompt injection with preference
  optimization.
\newblock In {\em Proceedings of the 2025 ACM SIGSAC Conference on Computer and
  Communications Security}, pages 2833--2847, 2025.
\newblock URL: \url{https://doi.org/10.1145/3719027.3744836}.

\bibitem{chen2024agentpoison}
Zhaorun Chen, Zhen Xiang, Chaowei Xiao, Dawn Song, and Bo~Li.
\newblock Agentpoison: Red-teaming llm agents via poisoning memory or knowledge
  bases.
\newblock In {\em Advances in Neural Information Processing Systems},
  volume~37, pages 130185--130213, 2024.
\newblock URL: \url{https://arxiv.org/abs/2407.12784}.

\bibitem{christodorescu2026foundations}
Mihai Christodorescu, Earlence Fernandes, Ashish Hooda, Somesh Jha, Johann
  Rehberger, Kamalika Chaudhuri, Xiaohan Fu, Khawaja Shams, Guy Amir, Jihye
  Choi, et~al.
\newblock Systems security foundations for agentic computing.
\newblock {\em arXiv preprint arXiv:2512.01295}, 2025.
\newblock URL: \url{https://arxiv.org/abs/2512.01295}.

\bibitem{coll2025digitalme}
Llu{\'\i}s~C Coll, Martin~W Lauer-Schmaltz, Philip Cash, John~P Hansen, and
  Anja Maier.
\newblock Towards the" digital me": a vision of authentic conversational agents
  powered by personal human digital twins.
\newblock {\em arXiv preprint arXiv:2506.23826}, 2025.
\newblock URL: \url{https://arxiv.org/abs/2506.23826}.

\bibitem{costa2025fides}
Manuel Costa, Boris K{\"o}pf, Aashish Kolluri, Andrew Paverd, Mark Russinovich,
  Ahmed Salem, Shruti Tople, Lukas Wutschitz, and Santiago
  Zanella-B{\'e}guelin.
\newblock Securing ai agents with information-flow control.
\newblock {\em arXiv preprint arXiv:2505.23643}, 2025.
\newblock URL: \url{https://arxiv.org/abs/2505.23643}.

\bibitem{debenedetti2025camel}
Edoardo Debenedetti, Ilia Shumailov, Tianqi Fan, Jamie Hayes, Nicholas Carlini,
  Daniel Fabian, Christoph Kern, Chongyang Shi, Andreas Terzis, and Florian
  Tram{\`e}r.
\newblock Defeating prompt injections by design.
\newblock {\em arXiv preprint arXiv:2503.18813}, 2025.
\newblock URL: \url{https://arxiv.org/abs/2503.18813}.

\bibitem{debenedetti2024agentdojo}
Edoardo Debenedetti, Jie Zhang, Mislav Balunovic, Luca Beurer-Kellner, Marc
  Fischer, and Florian Tram{\`e}r.
\newblock Agentdojo: A dynamic environment to evaluate prompt injection attacks
  and defenses for llm agents.
\newblock In {\em Advances in neural information processing systems},
  volume~37, pages 82895--82920, 2024.
\newblock URL: \url{https://arxiv.org/abs/2406.13352}.

\bibitem{dong2026memoryinjection}
Shen Dong, Shaochen Xu, Pengfei He, Yige Li, Jiliang Tang, Tianming Liu, Hui
  Liu, and Zhen Xiang.
\newblock Memory injection attacks on llm agents via query-only interaction.
\newblock {\em Advances in Neural Information Processing Systems},
  38:46697--46731, 2026.
\newblock URL: \url{https://arxiv.org/abs/2503.03704}.

\bibitem{du2025twinvoice}
Bangde Du, Minghao Guo, Songming He, Ziyi Ye, Xi~Zhu, Weihang Su, Shuqi Zhu,
  Yujia Zhou, Yongfeng Zhang, Qingyao Ai, et~al.
\newblock Twinvoice: A multi-dimensional benchmark towards digital twins via
  llm persona simulation.
\newblock In {\em Findings of the Association for Computational Linguistics:
  ACL 2026}, pages 19604--19628, 2026.
\newblock URL: \url{https://arxiv.org/abs/2510.25536}.

\bibitem{felt2011demystified}
Adrienne~Porter Felt, Erika Chin, Steve Hanna, Dawn Song, and David Wagner.
\newblock Android permissions demystified.
\newblock In {\em Proceedings of the 18th ACM conference on Computer and
  communications security}, pages 627--638, 2011.
\newblock URL: \url{https://doi.org/10.1145/2046707.2046779}.

\bibitem{felt2012attention}
Adrienne~Porter Felt, Elizabeth Ha, Serge Egelman, Ariel Haney, Erika Chin, and
  David Wagner.
\newblock Android permissions: User attention, comprehension, and behavior.
\newblock In {\em Proceedings of the eighth symposium on usable privacy and
  security}, pages 1--14, 2012.
\newblock URL: \url{https://doi.org/10.1145/2335356.2335360}.

\bibitem{geng2025controlillusion}
Yilin Geng, Haonan Li, Honglin Mu, Xudong Han, Timothy Baldwin, Omri Abend,
  Eduard Hovy, and Lea Frermann.
\newblock Control illusion: The failure of instruction hierarchies in large
  language models.
\newblock In {\em Proceedings of the AAAI Conference on Artificial
  Intelligence}, volume~40, pages 30816--30824, 2026.
\newblock URL: \url{https://arxiv.org/abs/2502.15851}.

\bibitem{gong2026planguard}
Guangyu Gong and Zizhuang Deng.
\newblock Planguard: Defending agents against indirect prompt injection via
  planning-based consistency verification.
\newblock {\em arXiv preprint arXiv:2604.10134}, 2026.
\newblock URL: \url{https://arxiv.org/abs/2604.10134}.

\bibitem{geminicli}
Google.
\newblock {Gemini CLI}.
\newblock GitHub repository, commit
  \texttt{571851b1077a51cef757146ce13f9da887326bec}, 2026.
\newblock Accessed 2026-08-19.
\newblock URL: \url{https://github.com/google-gemini/gemini-cli}.

\bibitem{abdelnabi2023ipi}
Kai Greshake, Sahar Abdelnabi, Shailesh Mishra, Christoph Endres, Thorsten
  Holz, and Mario Fritz.
\newblock Not what you've signed up for: Compromising real-world llm-integrated
  applications with indirect prompt injection.
\newblock In {\em Proceedings of the 16th ACM workshop on artificial
  intelligence and security}, pages 79--90, 2023.
\newblock URL: \url{https://doi.org/10.1145/3605764.3623985}.

\bibitem{cisco2026memory}
Idan Habler and Amy Chang.
\newblock {Identifying and Remediating a Persistent Memory Compromise in Claude
  Code}.
\newblock Cisco Blogs, 2026.
\newblock Published 2026-04-01; accessed 2026-08-19.
\newblock URL:
  \url{https://blogs.cisco.com/ai/identifying-and-remediating-a-persistent-memory-compromise-in-claude-code}.

\bibitem{he2026memoryarena}
Zexue He, Yu~Wang, Churan Zhi, Yuanzhe Hu, Tzu-Ping Chen, Lang Yin, Ze~Chen,
  Tong~Arthur Wu, Siru Ouyang, Zihan Wang, et~al.
\newblock Memoryarena: Benchmarking agent memory in interdependent
  multi-session agentic tasks.
\newblock {\em arXiv preprint arXiv:2602.16313}, 2026.
\newblock URL: \url{https://arxiv.org/abs/2602.16313}.

\bibitem{hines2024spotlighting}
Keegan Hines, Gary Lopez, Matthew Hall, Federico Zarfati, Yonatan Zunger, and
  Emre Kiciman.
\newblock Defending against indirect prompt injection attacks with
  spotlighting.
\newblock {\em arXiv preprint arXiv:2403.14720}, 2024.
\newblock URL: \url{https://arxiv.org/abs/2403.14720}.

\bibitem{howell2026agent}
Jason Howell.
\newblock {The Agent That Deleted A Company in 9 Seconds}.
\newblock Experience Tech, 2026.
\newblock Published 2026-04-29; accessed 2026-08-19.
\newblock URL:
  \url{https://jasonhowell.substack.com/p/the-agent-that-deleted-a-company}.

\bibitem{hu2026saber}
Qi~Hu, Yifeng Tang, Qinghua Wang, Lanyang Zhao, Pengji Zhang, Yuhao Qing, Xin
  Yao, Dong Huang, Lin Zhang, and Zhuoran Ji.
\newblock Saber: Benchmarking operational safety of llm coding agents in
  stateful project workspaces.
\newblock {\em arXiv preprint arXiv:2606.01317}, 2026.
\newblock URL: \url{https://arxiv.org/abs/2606.01317}.

\bibitem{anthropic2026automode}
John Hughes.
\newblock {How we built Claude Code auto mode: a safer way to skip
  permissions}.
\newblock Anthropic engineering blog, 2026.
\newblock Accessed 2026-08-19.
\newblock URL:
  \url{https://www.anthropic.com/engineering/claude-code-auto-mode}.

\bibitem{ji2026permissiongate}
Zimo Ji, Zongjie Li, Wenyuan Jiang, Yudong Gao, and Shuai Wang.
\newblock Measuring the permission gate: A stress-test evaluation of claude
  code's auto mode.
\newblock {\em arXiv preprint arXiv:2604.04978}, 2026.
\newblock URL: \url{https://arxiv.org/abs/2604.04978}.

\bibitem{kariyappa2026air}
Sanjay Kariyappa and G~Edward Suh.
\newblock Stronger enforcement of instruction hierarchy via augmented
  intermediate representations.
\newblock {\em arXiv preprint arXiv:2505.18907}, 2025.
\newblock URL: \url{https://arxiv.org/abs/2505.18907}.

\bibitem{li2026ace}
Evan Li, Tushin Mallick, Evan Rose, William Robertson, Alina Oprea, and
  Cristina Nita{-}Rotaru.
\newblock {ACE: A Security Architecture for LLM-Integrated App Systems}.
\newblock In {\em Network and Distributed System Security Symposium ({NDSS})},
  2026.
\newblock \href {https://doi.org/10.14722/ndss.2026.230352}
  {\path{doi:10.14722/ndss.2026.230352}}.

\bibitem{li2025drift}
Hao Li, Xiaogeng Liu, CHIU Chun, Dianqi Li, Ning Zhang, and Chaowei Xiao.
\newblock Drift: Dynamic rule-based defense with injection isolation for
  securing llm agents.
\newblock In {\em Advances in Neural Information Processing Systems},
  volume~38, pages 83262--83290, 2026.
\newblock URL: \url{https://arxiv.org/abs/2506.12104}.

\bibitem{li2026agentdyn}
Hao Li, Ruoyao Wen, Shanghao Shi, Ning Zhang, Yevgeniy Vorobeychik, and Chaowei
  Xiao.
\newblock Agentdyn: Are your agent security defenses deployable in real-world
  dynamic environments?
\newblock {\em arXiv preprint arXiv:2602.03117}, 2026.
\newblock URL: \url{https://arxiv.org/abs/2602.03117}.

\bibitem{li2025behaviorchain}
Rui Li, Heming Xia, Xinfeng Yuan, Qingxiu Dong, Lei Sha, Wenjie Li, and Zhifang
  Sui.
\newblock How far are llms from being our digital twins? a benchmark for
  persona-based behavior chain simulation.
\newblock In {\em Findings of the Association for Computational Linguistics:
  ACL 2025}, pages 15738--15763, 2025.
\newblock URL: \url{https://doi.org/10.18653/v1/2025.findings-acl.813}.

\bibitem{li2026conleash}
Ying Li, Yanju Chen, Peiran Wang, Issac Khabra, Faysal~Hossain Shezan, Yu~Feng,
  and Yuan Tian.
\newblock Options, not clicks: Lattice refinement for consent-driven mcp
  authorization.
\newblock {\em arXiv preprint arXiv:2605.11360}, 2026.
\newblock URL: \url{https://arxiv.org/abs/2605.11360}.

\bibitem{liu2024formalizing}
Yupei Liu, Yuqi Jia, Runpeng Geng, Jinyuan Jia, and Neil~Zhenqiang Gong.
\newblock Formalizing and benchmarking prompt injection attacks and defenses.
\newblock In {\em 33rd USENIX Security Symposium (USENIX Security 24)}, pages
  1831--1847, 2024.
\newblock URL: \url{https://arxiv.org/abs/2310.12815}.

\bibitem{maharana2024locomo}
Adyasha Maharana, Dong-Ho Lee, Sergey Tulyakov, Mohit Bansal, Francesco
  Barbieri, and Yuwei Fang.
\newblock Evaluating very long-term conversational memory of llm agents.
\newblock In {\em Proceedings of the 62nd Annual Meeting of the Association for
  Computational Linguistics (Volume 1: Long Papers)}, pages 13851--13870, 2024.
\newblock URL: \url{https://doi.org/10.18653/v1/2024.acl-long.747}.

\bibitem{maloyan2026codingassistants}
Narek Maloyan and Dmitry Namiot.
\newblock Prompt injection attacks on agentic coding assistants: A systematic
  analysis of vulnerabilities in skills, tools, and protocol ecosystems.
\newblock {\em International Journal of Open Information Technologies},
  14(2):1--10, 2026.
\newblock URL: \url{https://arxiv.org/abs/2601.17548}.

\bibitem{nanoclaw}
{NanoCo AI}.
\newblock {NanoClaw}.
\newblock GitHub repository, 2026.
\newblock Accessed 2026-08-20.
\newblock URL: \url{https://github.com/nanocoai/nanoclaw}.

\bibitem{openai2026huggingface}
{OpenAI}.
\newblock {OpenAI and Hugging Face Partner to Address Security Incident During
  Model Evaluation}.
\newblock OpenAI, 2026.
\newblock Accessed 2026-08-19.
\newblock URL:
  \url{https://openai.com/index/hugging-face-model-evaluation-security-incident/}.

\bibitem{codex}
{OpenAI}.
\newblock {OpenAI Codex}, 2026.
\newblock Accessed 2026-08-20.
\newblock URL: \url{https://github.com/openai/codex}.

\bibitem{openclaw}
{OpenClaw}.
\newblock {OpenClaw}, 2026.
\newblock Accessed 2026-08-20.
\newblock URL: \url{https://github.com/openclaw/openclaw}.

\bibitem{packer2024memgpt}
Charles Packer, Sarah Wooders, Kevin Lin, Vivian Fang, Shishir~G Patil, Ion
  Stoica, and Joseph~E Gonzalez.
\newblock Memgpt: Towards llms as operating systems.
\newblock {\em arXiv preprint arXiv:2310.08560}, 2023.
\newblock URL: \url{https://arxiv.org/abs/2310.08560}.

\bibitem{park2023generativeagents}
Joon~Sung Park, Joseph O'Brien, Carrie~Jun Cai, Meredith~Ringel Morris, Percy
  Liang, and Michael~S Bernstein.
\newblock Generative agents: Interactive simulacra of human behavior.
\newblock In {\em Proceedings of the 36th annual acm symposium on user
  interface software and technology}, pages 1--22, 2023.
\newblock URL: \url{https://doi.org/10.1145/3586183.3606763}.

\bibitem{patlan2025contextmanipulation}
Atharv~Singh Patlan, Ashwin Hebbar, Pramod Viswanath, and Prateek Mittal.
\newblock Context manipulation attacks: Web agents are susceptible to corrupted
  memory.
\newblock {\em arXiv preprint arXiv:2506.17318}, 2025.
\newblock URL: \url{https://arxiv.org/abs/2506.17318}.

\bibitem{perez2022ignore}
F{\'a}bio Perez and Ian Ribeiro.
\newblock Ignore previous prompt: Attack techniques for language models.
\newblock {\em arXiv preprint arXiv:2211.09527}, 2022.
\newblock URL: \url{https://arxiv.org/abs/2211.09527}.

\bibitem{piet2026planthenexecute}
Julien Piet, Annabella Chow, Yiwei Hou, Muxi Lyu, Sylvie Venuto, Jinhao Zhu,
  Raluca~Ada Popa, and David Wagner.
\newblock Web agents should adopt the plan-then-execute paradigm.
\newblock {\em arXiv preprint arXiv:2605.14290}, 2026.
\newblock URL: \url{https://arxiv.org/abs/2605.14290}.

\bibitem{pulipaka2026sleeper}
Sidharth Pulipaka, Stanislau Hlebik, Leonidas Raghav, Sahar Abdelnabi, Vyas
  Raina, Ivaxi Sheth, and Mario Fritz.
\newblock Hidden in memory: Sleeper memory poisoning in llm agents.
\newblock {\em arXiv preprint arXiv:2605.15338}, 2026.
\newblock URL: \url{https://arxiv.org/abs/2605.15338}.

\bibitem{qin2026airguard}
Suliu Qin, Haomin Zhuang, Yujun Zhou, Yufei Han, and Xiangliang Zhang.
\newblock Airguard: Guarding agent actions with runtime authority control.
\newblock {\em arXiv preprint arXiv:2605.28914}, 2026.
\newblock URL: \url{https://arxiv.org/abs/2605.28914}.

\bibitem{saltzer1975protection}
Jerome~H. Saltzer and Michael~D. Schroeder.
\newblock {The Protection of Information in Computer Systems}.
\newblock {\em Proceedings of the {IEEE}}, 63(9):1278--1308, 1975.
\newblock \href {https://doi.org/10.1109/PROC.1975.9939}
  {\path{doi:10.1109/PROC.1975.9939}}.

\bibitem{santosgrueiro2026portico}
Igor Santos-Grueiro.
\newblock Lingering authority: Revocable resource-and-effect capabilities for
  coding agents.
\newblock {\em arXiv preprint arXiv:2606.22504}, 2026.
\newblock URL: \url{https://arxiv.org/abs/2606.22504}.

\bibitem{shi2025progent}
Tianneng Shi, Jingxuan He, Zhun Wang, Hongwei Li, Linyu Wu, Wenbo Guo, and Dawn
  Song.
\newblock Progent: Securing ai agents with privilege control.
\newblock {\em arXiv preprint arXiv:2504.11703}, 2025.
\newblock URL: \url{https://arxiv.org/abs/2504.11703}.

\bibitem{srivastava2025memorygraft}
Saksham~Sahai Srivastava and Haoyu He.
\newblock Memorygraft: Persistent compromise of llm agents via poisoned
  experience retrieval.
\newblock {\em arXiv preprint arXiv:2512.16962}, 2025.
\newblock URL: \url{https://arxiv.org/abs/2512.16962}.

\bibitem{guardian2026database}
{The Guardian}.
\newblock {Claude-powered AI agent’s confession after deleting a firm’s
  entire database: ‘I violated every principle I was given’}.
\newblock The Guardian, 2026.
\newblock Published 2026-04-29; accessed 2026-08-19.
\newblock URL:
  \url{https://www.theguardian.com/technology/2026/apr/29/claude-ai-deletes-firm-database}.

\bibitem{tsai2025conseca}
Lillian Tsai and Eugene Bagdasarian.
\newblock Contextual agent security: A policy for every purpose.
\newblock In {\em Proceedings of the 2025 Workshop on Hot Topics in Operating
  Systems}, pages 8--17, 2025.
\newblock URL: \url{https://doi.org/10.1145/3713082.3730378}.

\bibitem{wallace2024hierarchy}
Eric Wallace, Kai Xiao, Reimar Leike, Lilian Weng, Johannes Heidecke, and Alex
  Beutel.
\newblock The instruction hierarchy: Training llms to prioritize privileged
  instructions.
\newblock {\em arXiv preprint arXiv:2404.13208}, 2024.
\newblock URL: \url{https://arxiv.org/abs/2404.13208}.

\bibitem{wang2026landscape}
Peiran Wang, Xinfeng Li, Chong Xiang, Jinghuai Zhang, Ying Li, Lixia Zhang,
  Xiaofeng Wang, and Yuan Tian.
\newblock The landscape of prompt injection threats in llm agents: From
  taxonomy to analysis.
\newblock {\em arXiv preprint arXiv:2602.10453}, 2026.
\newblock URL: \url{https://arxiv.org/abs/2602.10453}.

\bibitem{wang2026pasb}
Yuhang Wang, Feiming Xu, Zheng Lin, Guangyu He, Yuzhe Huang, Haichang Gao,
  Zhenxing Niu, Shiguo Lian, and Zhaoxiang Liu.
\newblock From assistant to double agent: Formalizing and benchmarking attacks
  on openclaw for personalized local ai agent.
\newblock {\em arXiv preprint arXiv:2602.08412}, 2026.
\newblock URL: \url{https://arxiv.org/abs/2602.08412}.

\bibitem{wei2025amemguard}
Qianshan Wei, Tengchao Yang, Yaochen Wang, Xinfeng Li, Lijun Li, Zhenfei Yin,
  Yi~Zhan, Thorsten Holz, Zhiqiang Lin, and XiaoFeng Wang.
\newblock A-memguard: A proactive defense framework for llm-based agent memory.
\newblock {\em arXiv preprint arXiv:2510.02373}, 2025.
\newblock URL: \url{https://arxiv.org/abs/2510.02373}.

\bibitem{westhausser2025personalized}
Rebecca Westh{\"a}u{\ss}er, Wolfgang Minker, and Sebatian Zepf.
\newblock Enabling personalized long-term interactions in llm-based agents
  through persistent memory and user profiles.
\newblock {\em arXiv preprint arXiv:2510.07925}, 2025.
\newblock URL: \url{https://arxiv.org/abs/2510.07925}.

\bibitem{wu2025isolategpt}
Yuhao Wu, Franziska Roesner, Tadayoshi Kohno, Ning Zhang, and Umar Iqbal.
\newblock {IsolateGPT: An Execution Isolation Architecture for LLM-Based
  Agentic Systems}.
\newblock In {\em Network and Distributed System Security Symposium ({NDSS})},
  2025.
\newblock URL: \url{https://arxiv.org/abs/2403.04960}.

\bibitem{wu2026permissions}
Yuhao Wu, Ke~Yang, Franziska Roesner, Tadayoshi Kohno, Ning Zhang, and Umar
  Iqbal.
\newblock Towards automating data access permissions in ai agents.
\newblock In {\em 2026 IEEE Symposium on Security and Privacy (SP)}, pages
  2609--2627. IEEE, 2026.
\newblock URL: \url{https://doi.org/10.1109/SP63933.2026.00018}.

\bibitem{xu2026memoryctrlflow}
Zhenlin Xu, Xiaogang Zhu, Yu~Yao, Minhui Xue, and Yiliao Song.
\newblock From storage to steering: Memory control flow attacks on llm agents.
\newblock {\em arXiv preprint arXiv:2603.15125}, 2026.
\newblock URL: \url{https://arxiv.org/abs/2603.15125}.

\bibitem{yan2026authbench}
Zheng Yan, Jingxiang Weng, Charles Chen, Dengyun Peng, Ethan Qin, Jiannan Guan,
  Jinhao Liu, Qiming Yu, Yixin Yuan, Fanqing Meng, et~al.
\newblock Do coding agents understand least-privilege authorization?
\newblock {\em arXiv preprint arXiv:2605.14859}, 2026.
\newblock URL: \url{https://arxiv.org/abs/2605.14859}.

\bibitem{yang2024llmtwinhitl}
Hanqing Yang, Marie Siew, and Carlee Joe-Wong.
\newblock An llm-based digital twin for optimizing human-in-the loop systems.
\newblock In {\em 2024 IEEE International Workshop on Foundation Models for
  Cyber-Physical Systems \& Internet of Things (FMSys)}, pages 26--31. IEEE,
  2024.
\newblock URL: \url{https://arxiv.org/abs/2403.16809}.

\bibitem{yang2026zombie}
Xianglin Yang, Yufei He, Shuo Ji, Bryan Hooi, and Jin~Song Dong.
\newblock Zombie agents: Persistent control of self-evolving llm agents via
  self-reinforcing injections.
\newblock {\em arXiv preprint arXiv:2602.15654}, 2026.
\newblock URL: \url{https://arxiv.org/abs/2602.15654}.

\bibitem{ye2026roleconfusion}
Charles Ye, Jasmine Cui, and Dylan Hadfield-Menell.
\newblock Prompt injection as role confusion.
\newblock {\em arXiv preprint arXiv:2603.12277}, 2026.
\newblock URL: \url{https://arxiv.org/abs/2603.12277}.

\bibitem{zhan2024injecagent}
Qiusi Zhan, Zhixiang Liang, Zifan Ying, and Daniel Kang.
\newblock Injecagent: Benchmarking indirect prompt injections in
  tool-integrated large language model agents.
\newblock In {\em Findings of the Association for Computational Linguistics:
  ACL 2024}, pages 10471--10506, 2024.
\newblock URL: \url{https://doi.org/10.18653/v1/2024.findings-acl.624}.

\bibitem{zhang2025asb}
Hanrong Zhang, Jingyuan Huang, Kai Mei, Yifei Yao, Zhenting Wang, Chenlu Zhan,
  Hongwei Wang, and Yongfeng Zhang.
\newblock Agent security bench (asb): Formalizing and benchmarking attacks and
  defenses in llm-based agents.
\newblock In {\em International Conference on Learning Representations}, volume
  2025, pages 35331--35366, 2025.
\newblock URL: \url{https://arxiv.org/abs/2410.02644}.

\bibitem{zhang2024memorysurvey}
Zeyu Zhang, Quanyu Dai, Xiaohe Bo, Chen Ma, Rui Li, Xu~Chen, Jieming Zhu,
  Zhenhua Dong, and Ji-Rong Wen.
\newblock A survey on the memory mechanism of large language model-based
  agents.
\newblock {\em ACM Transactions on Information Systems}, 43(6):1--47, 2025.
\newblock URL: \url{https://arxiv.org/abs/2404.13501}.

\bibitem{zhao2026amabench}
Yujie Zhao, Boqin Yuan, Junbo Huang, Haocheng Yuan, Zhongming Yu, Haozhou Xu,
  Lanxiang Hu, Abhilash Shankarampeta, Zimeng Huang, Wentao Ni, et~al.
\newblock Ama-bench: Evaluating long-horizon memory for agentic applications.
\newblock {\em arXiv preprint arXiv:2602.22769}, 2026.
\newblock URL: \url{https://arxiv.org/abs/2602.22769}.

\bibitem{zheng2026actplane}
Yusheng Zheng, Tianyuan Wu, Quanzhi Fu, Tong Yu, Wenan Mao, Tao Ma, Dan
  Williams, Wei Wang, and Andi Quinn.
\newblock Actplane: Programmable os-level policy enforcement for agent
  harnesses.
\newblock {\em arXiv preprint arXiv:2606.25189}, 2026.
\newblock URL: \url{https://arxiv.org/abs/2606.25189}.

\bibitem{zhu2025miniscope}
Jinhao Zhu, Kevin Tseng, Gil Vernik, Xiao Huang, Shishir~G Patil, Vivian Fang,
  and Raluca~Ada Popa.
\newblock Miniscope: A least privilege framework for authorizing tool calling
  agents.
\newblock {\em arXiv preprint arXiv:2512.11147}, 2025.
\newblock URL: \url{https://arxiv.org/abs/2512.11147}.

\bibitem{zou2026poisononce}
Wei Zou, Mingwen Dong, Miguel~Romero Calvo, Shuaichen Chang, Jiang Guo, Dongkyu
  Lee, Xing Niu, Xiaofei Ma, Yanjun Qi, and Jiarong Jiang.
\newblock Poison once, exploit forever: Environment-injected memory poisoning
  attacks on web agents.
\newblock {\em arXiv preprint arXiv:2604.02623}, 2026.
\newblock URL: \url{https://arxiv.org/abs/2604.02623}.

\end{thebibliography}

\end{document}